\documentclass[10.5pt,letter]{article}

\usepackage{amsmath, amsthm, amssymb}
\allowdisplaybreaks 

\usepackage{amssymb,epsfig}

\usepackage{hyperref}
\hypersetup{bookmarks=true}

\makeatletter

\gdef\@punct{.\ \ }  
\def\@sect#1#2#3#4#5#6[#7]#8{%
  \ifnum #2>\c@secnumdepth
     \def\@svsec{}
  \else
     \refstepcounter{#1}\edef\@svsec{%
     \ifnum #2>0{{\csname the#1\endcsname}}.\fi%
    \hskip .5em}
  \fi
  \@tempskipa #5\relax
  \ifdim \@tempskipa>\z@
     \begingroup #6\relax
       \@hangfrom{\hskip #3\relax\@svsec}{\interlinepenalty \@M #8\par}
     \endgroup
     \csname #1mark\endcsname{#7}
     \addcontentsline{toc}{#1}{\ifnum #2>\c@secnumdepth\else
          \protect\numberline{\csname the#1\endcsname}\fi#7}
  \else
     \def\@svsechd{#6\hskip #3\@svsec #8\@punct\csname
#1mark\endcsname{#7}
     \addcontentsline{toc}{#1}{\ifnum #2>\c@secnumdepth \else
          \protect\numberline{\csname the#1\endcsname}\fi#7}}
  \fi
  \@xsect{#5}}

\def\@ssect#1#2#3#4#5{\@tempskipa #3\relax
  \ifdim \@tempskipa>\z@
     \begingroup #4\@hangfrom{\hskip #1}{\interlinepenalty \@M
#5\par}\endgroup
  \else \def\@svsechd{#4\hskip #1\relax #5\@punct}\fi
  \@xsect{#3}}

\def\qed{\hskip 3pt \hbox{\vrule width4pt depth2pt height6pt}}

\newtheorem{Lemma}{Lemma}

\newtheorem{Theorem}[Lemma]{Theorem}
\newtheorem{Proposition}[Lemma]{Proposition}
\newtheorem{Corollary}[Lemma]{Corollary}
\newtheorem{Definition}[Lemma]{Definition}

\usepackage{calc}
\usepackage{tikz}
\usetikzlibrary{decorations.markings}
\tikzstyle{vertex}=[circle, draw, inner sep=0pt, minimum size=6pt]

\tikzset{->-/.style={decoration={
markings,
mark=at position #1 with {\arrow{>}}},postaction={decorate}}}

\usetikzlibrary{intersections}

\usepackage[ruled, vlined, linesnumbered]{algorithm2e}

\usepackage{caption}
\usepackage{mathptmx}
\usepackage[per-mode=symbol]{siunitx}
\usepackage{bm}
\usepackage{booktabs}

\begin{document}

\title{The Structure of Hypergraphs Arising in Cellular Mobile Communication Systems}

\author{Ashwin~Ganesan%
  \thanks{The author was formerly with ATLAS SkillTech University, Tower 1, Equinox Business Park, Ambedkar Nagar, Kurla West, Kurla, Mumbai 400070, Maharashtra, India.  E-mail: \texttt{ashwin.ganesan@gmail.com}. \\ This paper was presented in part at the International Conference on Distributed Computing and Networking (ICDCN), IIT Kharagpur, India, January 2023 \cite{Ganesan:ICDCN:23}.}
}

\date{}

\maketitle


\begin{abstract}
\noindent  An assumption that researchers have often used to model interference in a wireless network is the unit disk graph model.  While many theoretical results and performance guarantees have been obtained under this model, an open research direction is to extend these results to hypergraph interference models.  Motivated by recent results that the worst-case performance of the distributed maximal scheduling algorithm is characterized by the interference degree of the hypergraph, in the present work we investigate properties of the interference degree of the hypergraph and the structure of hypergraphs arising from physical constraints.  We show that the problem of computing the interference degree of a hypergraph is NP-hard and we prove some properties and results concerning this hypergraph invariant.  We investigate which hypergraphs are realizable, i.e. which hypergraphs arise in practice, based on physical constraints, as the interference model of a wireless network.  In particular, a question that arises naturally is: what is the maximal value of $r$ such that the hypergraph $K_{1,r}$ is realizable?  We determine this quantity for various integral and nonintegral values of the path loss exponent of signal propagation.  We also investigate hypergraphs generated by line networks.

\end{abstract}

\tableofcontents


\section{Introduction}

Combinatorial interference models for wireless communication networks such as conflict graphs and conflict hypergraphs have been widely studied in the literature, and the performance of various protocols and distributed algorithms for resource allocation and scheduling have been investigated using these models.  The combinatorial objects that arise in this context due to geometric, graph-theoretic, or physical constraints have additional structure that can be exploited to design efficient algorithms with near-optimal performance for various problems. Physical constraints can include the requirement that the SINR (Signal-to-Interference-plus-Noise Ratio) exceed a certain threshold.  For example, unit disk graphs have additional structure - the interference degree of a unit disk graph (defined below) is at most $5$ - and this implies that the performance of the distributed maximal scheduling algorithm is a factor of at most $5$ away from optimal.  An open problem in the literature that we consider in the present work is to extend the results in the literature on unit disk graphs to hypergraph models.

When entities in a network compete for resources, the competition can be modeled by a \emph{conflict graph}, whose vertices correspond to the entities and whose edges correspond to pairs of vertices that cannot be assigned the same resource.  Thus, many resource allocation and scheduling problems are variations of the problem of coloring the conflict graph \cite{Chaitin:1982} \cite{Hale:1980}.  In a wireless network, nodes which are in the same vicinity interfere with each other and correspond to adjacent vertices in the conflict graph \cite{Jain:Padhye:etal:03}.  Given the conflict graph of a wireless network and the fraction of each unit of time that each node demands to be active, the \emph{admission control problem} is to determine whether the set of demands is feasible. The optimal, centralized version of the admission control problem is NP-hard \cite{Ganesan:2010}.  

A distributed algorithm for the admission control problem that relies only on localized information and has low processing complexity is maximal scheduling \cite{Wu:Srikant:2006}, also known as rate constraints \cite{Hamdaoui:Ramanathan:03} \cite{Hamdaoui:Ramanathan:05} or row constraints \cite{Gupta:Musacchio:Walrand:07}.  The \emph{interference degree} of a graph is the maximum number of vertices that are neighbors of a vertex and that are pairwise nonadjacent.   In \cite{Chaporkar:Kar:Luo:Sarkar:2008} \cite{Ganesan:2009} \cite{Ganesan:2010} it was shown that in the worst case, the performance of the maximal scheduling algorithm is away from optimal by a factor equal to the interference degree of the conflict graph. 

In some real world problems, the conflict graph has additional structure, based on geometric or other constraints, which can be exploited to give efficient algorithms whose suboptimality is bounded by a fixed constant.  For example, consider a wireless network whose nodes are equipped with omnidirectional antennas and have the same transmission range.  Suppose two nodes interfere with each other whenever unit disks centered at their locations intersect \cite{Huson:Sen:1995}.  The resulting conflict graph is called a unit disk graph.  The interference degree of a unit disk graph is at most $5$ \cite{MaratheEtAl:1995}.  By the results in the preceding paragraph, the maximal scheduling algorithm is a factor of  $5$ away from optimal in the worst case for unit disk graphs \cite{Ganesan:2009} \cite{Ganesan:WN:2014}. An open problem in the literature (cf. \cite[p. 1333]{Ganesan:WN:2014}) is to extend the results in the literature on unit disk graphs to hypergraph models. 

Hypergraph interference models and physical (SINR) models  capture certain complexities which cannot be captured by graphs and hence  achieve better performance  \cite{McEliece:Sivarajan:1991} \cite{McEliece:Sivarajan:1994} \cite{Sarkar:Sivarajan:1998} \cite{Zhang:Song:Han:2016} \cite{Halldorsson:Wattenhofer:2019}.  For example, in a wireless network with three nodes, it is possible for the interference to be such that any two nodes can be simultaneously active, and if all three nodes are simultaneously active then the interference is intolerable.  A subset of nodes that causes intolerance interference is called a forbidden set.  The interference can be modeled by a hypergraph, whose hyperedges are the minimal forbidden sets.  A maximal scheduling algorithm for wireless networks under the hypergraph interference model was proposed in \cite{Li:Negi:2012}.  The interference degree of a hypergraph was defined in \cite{Ganesan:TIT:2021} and was shown to characterize the worst-case performance of the maximal scheduling algorithm.

In the present work, we investigate the interference degree of structured hypergraphs. The problem of computing the interference degree of a hypergraph is shown to be NP-hard, and the interference degree of certain structured hypergraphs is determined. We also investigate which hypergraphs are realizable, i.e. which hypergraphs arise as the interference model of a wireless network.  For various values of the path loss exponent, we determine the maximal value of $r$ such that the hypergraph $K_{1,r}$ is realizable.

The rest of this paper is organized as follows.  Section~\ref{sec:system:model} describes the system model; formal definitions of the unit disk graph model and hypergraph model are given, a distributed maximal scheduling algorithm from the literature is recalled, and it is explained that its worst-case performance is characterized by the interference degree of the hypergraph.  Section~\ref{sec:related:work} discusses related work.  In Section~\ref{sec:interference:degree:hyp}, we show that the problem of computing the interference degree of a hypergraph is NP-hard and determine the interference degree of some structured hypergraphs.  In Section~\ref{sec:preliminaries} we prove some preliminary results, which are used in Section~\ref{sec:alpha:4}, Section~\ref{sec:alpha:3}, Section~\ref{sec:alpha:2}, and Section~\ref{sec:alpha:1} to prove that certain hypergraphs are realizable or nonrealizable when the path loss exponent $\gamma=4,3,2$ and $1$, respectively.  In Section~\ref{sec:nonintegral:alpha}, these results are extended to nonintegral values of the  path loss exponent $\gamma$.  In Section~\ref{sec:line:networks} some results concerning the hypergraphs of line networks are obtained. Section~\ref{sec:concluding:remarks} contains concluding remarks. To facilitate understanding, we provide in Table~\ref{tab:symbols} a list of all symbols used along with their corresponding meanings and the section where they are first defined.

\begin{table*}[htbp]
	\centering
	\begin{tabular}{lll}
		\textbf{Symbol} & \textbf{Meaning} & \textbf{Section} \\
		\hline
		$S$ & A set of $\{s_1,\ldots,s_n\} \subseteq \mathbb{R}^2$ of stations in the Euclidean plane & \ref{sec:udg} \\
		$G=(V,E)$ & a simple, undirected graph & \ref{sec:udg} \\
		$\alpha(G)$ & independence number of the graph $G$ & \ref{sec:udg} \\
		$\Gamma(v)$ & the set of neighbors of vertex $v$ & \ref{sec:udg} \\
		$G[W]$ & subgraph of $G$ induced by $W$  & \ref{sec:udg} \\
		$\sigma(G)$ & interference degree of the graph $G$ & \ref{sec:udg} \\
		$K_{1,r}$ & the star graph, or the $2$-uniform star hypergraph & \ref{sec:udg} \\
		$\gamma$ & path loss exponent & \ref{sec:path:loss} \\
		$d(a,b)$ & Euclidean distance between points $a$ and $b$ & \ref{sec:path:loss} \\
		$H=(V, \mathcal{E})$ & hypergraph with vertex set $V$ and edge set $\mathcal{E}$ & \ref{sec:hyp:wireless} \\
		$\beta$ & SINR threshold for successful reception & \ref{sec:hyp:wireless} \\
		$\mathcal{A} = \langle S, \gamma, \beta \rangle$ & wireless network, with set of stations $S$, path loss exponent $\gamma$, and reception threshold  $\beta$ & \ref{sec:hyp:wireless}  \\
		$E(W,w)$ & energy or interference at $w$ due to transmission by stations in $W$ & \ref{sec:hyp:wireless} \\
		$\tau = (\tau(v): v \in V)$ & QoS requirement or demand vector & \ref{sec:maximal:interference:degree} \\
		$\mathcal{I}(H)$ & collection of all independent sets of the hypergraph $H$ & \ref{sec:maximal:interference:degree} \\
		$A = [a_{ij}]$ & vertex-independent set incidence matrix of a hypergraph & \ref{sec:maximal:interference:degree} \\
		$\chi_f(H,\tau)$ & fractional chromatic number of the weighted hypergraph $(H,\tau)$ & \ref{sec:maximal:interference:degree} \\
		$N(i)$ & set of neighbors in a hypergraph of vertex $i$ & \ref{sec:maximal:interference:degree} \\
		$\Delta_{ij}, \Delta_i', \Delta_i''$ & intermediate variables for interference degree calculation & \ref{sec:maximal:interference:degree} \\
		$B(H,\tau)$ & an upper bound on $\chi_f(H,\tau)$ & \ref{sec:maximal:interference:degree} \\
		$\eta(H)$ & worst-case performance of the maximal scheduling algorithm & \ref{sec:maximal:interference:degree} \\ 
		$\sigma(H)$ & interference degree of the hypergraph $H$ & \ref{sec:maximal:interference:degree} \\
		$[N]$ & the set $\{1,2,\ldots,N\}$ & \ref{sec:notation} \\
		$S^{(r)}$ & the collection of all $r$-subsets of $S$ & \ref{sec:notation} \\
		$K_n^{(r)}$ & the complete $r$-uniform hypergraph  & \ref{sec:notation} \\
		$\theta_i$ & angle between stations on a unit circle & \ref{sec:alpha:4} \\
		$\delta_i$ & translated version of angle between stations on a unit circle & \ref{sec:alpha:4} \\
		$f(\delta_1)$ & cumulative interference at a station due to transmissions by its two closest neighbors & \ref{sec:alpha:4} \\
		$g(\delta)$ & a translation of the function $f(\delta_1) $ & \ref{sec:alpha:4} \\
		$r(\gamma)$ & largest star hypergraph that is realizable for path loss exponent $\gamma$ & \ref{sec:nonintegral:alpha} \\
		$\mathcal{U}_k$ & a set of $k$ stations placed uniformly on the unit circle & \ref{sec:nonintegral:alpha} \\
		$u(\gamma)$ & largest $k$ such that $\mathcal{U}_k$ is feasible & \ref{sec:nonintegral:alpha} \\
		$P_n$ & the path graph on $n$ vertices, or the $2$-uniform path hypergraph on $n$ vertices & \ref{sec:line:networks} \\		
	\end{tabular}
	\caption{Notation index in the order of first occurrence in paper}
	\label{tab:symbols}
\end{table*}

\section{System Model} \label{sec:system:model}

\subsection{Unit Disk Graphs} \label{sec:udg}

Consider a wireless network where all nodes have the same transmission range and are equipped with omnidirectional antennas.  Then, the coverage area of each node is a disk centered at the location of that node.  Two nodes interfere with each other if and only if the disks centered at their locations intersect.  
Given a set $S = \{s_1,\ldots,s_n\} \subseteq \mathbb{R}^2$ of stations in the Euclidean plane, the \emph{unit disk graph generated by $S$} is the graph $G=(S,E)$ with vertex set $S$, and with two vertices $s_i, s_j \in S$ being adjacent in $G$ whenever the Euclidean distance $d(s_i,s_j)$ between the corresponding two stations is at most $1$.  A graph is called a \emph{unit disk graph} if it is generated by some set $S \subseteq \mathbb{R}^2$.  If $S \subseteq \mathbb{R}^2$ generates $G$, then we say $S$ is a \emph{realization} of $G$.   The unit disk graph assumption has been used often  by researchers in the sensor networks community to model the topology of wireless networks \cite{Huson:Sen:1995}.  

Given a graph $G=(V,E)$, a vertex $v \in V$ and a subset $W \subseteq V$, let $\alpha(G)$ denote the independence number of $G$, let $\Gamma(v)$ denote the set of neighbors in $G$ of vertex $v$, and let $G[W]$ denote the subgraph of $G$ induced by $W$.  The \emph{interference degree} of a graph $G$, denoted by $\sigma(G)$, is defined to be $\max_{v \in V} \alpha(G[\Gamma(v)])$ \cite{Ganesan:2010} \cite{Chaporkar:Kar:Luo:Sarkar:2008}.  Thus, $\sigma(G)=r$ if and only if $G$ contains the star $K_{1,r}$ as an induced subgraph and $G$ does not contain the star $K_{1,r+1}$ as an induced subgraph.

It can be shown that the star $K_{1,5}$ is a unit disk graph, and that the star $K_{1,6}$ is not a unit disk graph \cite{MaratheEtAl:1995}.  To see why $K_{1,6}$ is not a unit disk graph, observe that the maximum distance between any two points in a $60^\circ$ sector of a unit circle is $1$; hence, if six stations are placed in the unit ball, then the distance between some two stations is at most $1$.   It follows that the interference degree of a unit disk graph is at most $5$.  This implies that a certain greedy, distributed algorithm for admission control, called maximal scheduling (or row constraints) (cf. \cite{Gupta:Musacchio:Walrand:07} \cite{Hamdaoui:Ramanathan:05}), is a factor of at most $5$ away from optimal for unit disk graphs \cite{Ganesan:WN:2014}.

%

\subsection{Signal Propagation Model and Path Loss Exponents} \label{sec:path:loss}

The path loss exponent plays a pivotal role in characterizing the attenuation of a wireless signal as it traverses through the radio channel. This exponent, denoted here by $\gamma$, quantifies the rate at which the signal strength diminishes with increasing distance from the transmitter. Understanding the path loss exponent is crucial in designing and optimizing wireless communication systems, as it directly influences signal coverage, range, and overall system performance. The path loss exponent is particularly significant in the deployment of wireless networks, where variations in the propagation environment necessitate a nuanced consideration of signal attenuation for efficient communication and reliable connectivity.

While signal propagation effects exhibit complexity in real channels and vary across different environments, a simplified path loss model is often employed for the analysis of tradeoffs and system design. This model expresses the received signal strength as a function of distance. It is assumed that the power level is the same for all transmitters and that the signal strength degrades uniformly in all directions. In the scenario where $s_i$ acts as the transmitter, $s_j$ serves as the receiver, and $d(s_i,s_j)$ represents the distance between them, the received signal strength at $s_j$ is defined as $\frac{1}{d(s_i,s_j)^\gamma}$, with $\gamma$ denoting the path loss exponent. This model has  been used in the literature on hypergraph interference models (defined below), where $\gamma$ was taken to equal $4$ in \cite{McEliece:Sivarajan:1994} \cite{Sarkar:Sivarajan:1998} \cite{Sarkar:Sivarajan:VTC:2002}.  Typical values of the path loss exponent for various environments are presented in Table~\ref{tab:path:loss}; cf \cite{Friis:1946} \cite{Goldsmith:2020}.  In most real environments, the path loss exponent $\gamma > 1$, although $\gamma < 1$ in certain indoor settings.  In this paper, it is assumed throughout that $\gamma > 0$. 

\begin{table}[htbp]
	\centering
	\begin{tabular}{ll}
		\textbf{Environment} & \textbf{$\gamma$ range} \\
		\hline
		Urban macrocells & 3.7-6.5 \\
		Urban microcells & 2.7-3.5 \\
		Office building (same floor) & 1.6-3.5 \\
		Office building (multiple floors) & 2-6 \\
		Store & 1.8-2.2 \\
		Factory & 1.6-3.3 \\
		Home & 3
	\end{tabular}
	\caption{Typical path loss exponents. Table courtesy of \cite[Section 2.5]{Goldsmith:2020}.}
	\label{tab:path:loss}
\end{table}

\subsection{The Hypergraph of a Wireless Network} \label{sec:hyp:wireless}

A hypergraph is a pair $(V,\mathcal{E})$ where $V$ is a finite set, called the vertex set or ground set, and $\mathcal{E}$ is a collection of subsets of $V$.  The elements of $\mathcal{E}$ are called hyperedges (or edges).  Thus, hypergraphs generalize graphs in the sense that the size of a hyperedge $E \in \mathcal{E}$ is not restricted to be $2$.

Consider a wireless network $\mathcal{A} = \langle S, \gamma, \beta \rangle$, where $S = \{s_1, s_2, \ldots, s_n\}$ is a set of stations (or transceivers), $\gamma$ is the path-loss exponent, and $\beta$ is the reception threshold. The parameter $\beta$ is sometimes also called the SINR threshold for successful reception \cite{Sarkar:Sivarajan:1998} (though we ignore noise in the present model), and refers to the maximum interference that is tolerable. We shall identify the symbol $s_i$, used for a station, with its location in the Euclidean plane, so that $S \subseteq \mathbb{R}^2$.  The value of the path loss exponent $\gamma$ typically varies between $1.6$ and $6.5$ \cite{Goldsmith:2020}. 

For $W \subseteq S$, the energy (or interference) at $w \in W$ due to transmissions by the stations in $W - \{w\}$ is defined to be
$$E(W-\{w\}, w) = \sum_{s \in W - \{w\}} \frac{1}{d(s,w)^\gamma},$$ 
where $d(a,b)$ denotes the Euclidean distance between points $a$ and $b$. 

Given a wireless network $\mathcal{A} = \langle S, \gamma, \beta \rangle$, a subset $W \subseteq S$ is called a \emph{forbidden set} for the wireless network $\mathcal{A}$   if the interference is intolerable at some station in $W$ when all the other stations in $W$ are active.  In other words, a subset $W$ of stations is forbidden if there exists a station $w \in W$ such that
$$
E(W-\{w\},w) = \sum_{s \in W - \{w\}} \frac{1}{d(s,w)^\gamma} \ge \beta.
$$
A forbidden set is a \emph{minimal forbidden set} if it does not properly contain a forbidden set.  The hypergraph interference model is defined by taking the hyperedges to be the minimal forbidden sets: 

\begin{Definition} \label{def:hypergraph:generated:by:wireless}
Let $\mathcal{A} = \langle S, \gamma, \beta \rangle$ be a wireless network, where $S$ is a set of points in $\mathbb{R}^2$, $\gamma$ is the path loss exponent, and $\beta$ is the reception threshold.  The \emph{hypergraph generated by $\mathcal{A}$} is defined to be the hypergraph $H = (S, \mathcal{E})$, where $\mathcal{E}$ is the collection of all minimal forbidden sets.  
\end{Definition}

This additive model of interference and definition of the hypergraph as a collection of minimal forbidden subsets has been investigated in \cite{McEliece:Sivarajan:1991} \cite{McEliece:Sivarajan:1994} \cite{Sarkar:Sivarajan:1998}. If the wireless network $\mathcal{A}$ generates the hypergraph $H$, then $\mathcal{A}$ is called a \emph{realization} of $H$.   An \emph{independent set} (or \emph{feasible set}) of the hypergraph $H=(S,\mathcal{E})$ is a subset of $S$ that does not contain any hyperedges. Thus, an independent set corresponds to a subset of the stations that can be simultaneously active.  Clearly, a subset of an independent set of a hypergraph is  an independent set, and a superset of a forbidden set is a forbidden set.

\subsection{A Distributed Maximal Scheduling Algorithm and The Interference Degree of a Hypergraph} \label{sec:maximal:interference:degree}

In this subsection, we recall from the literature  
a certain distributed maximal scheduling algorithm \cite{Li:Negi:2012} and explain how its worst-case performance is characterized by the interference degree of the hypergraph \cite{Ganesan:TIT:2021}. 

Consider a network whose entities compete for some resource, and the competition is modeled by a conflict hypergraph $H=(V,\mathcal{E})$, where the vertex set $V$ represents the set of entities, and each hyperedge $E \in \mathcal{E}$ represents a minimal collection of entities that cannot all be assigned the  resource at the same time due to said conflict or competition.  Furthermore, each entity $v \in V$ specifies a fraction $\tau(v)$ of each unit of time that it would like to use the resource.  The objective is to determine whether the demand vector $\tau$ can be satisfied.  For example, in a wireless network, the entities are the wireless links (or stations), and links in the same vicinity contend for the shared wireless medium due to interference.  The quality-of-service (QoS) requirement is specified in terms of a minimum bandwidth requirement for each link, which translates into a fraction of each unit of time when the link should be active.  Each hyperedge represents a minimal forbidden set of links.  The \emph{admission control problem} is to determine if the demand vector is \emph{feasible}, i.e. whether all the demands can be satisfied.  It is preferable that this decision be made in a distributed manner, i.e. using only localized information.

The admission control problem can be reformulated as follows.  Let $H=(V, \mathcal{E})$ be a conflict hypergraph, with $V = \{v_1,\ldots,v_n\}$, and let $\tau = (\tau(v): v \in V)$ be a demand vector.  Let $\mathcal{I}(H) = \{I_1, \ldots, I_L\}$ be the collection of all independent sets of $H$.  Let $A = [a_{ij}]$ be the $0,1$-valued $N \times L$ vertex-independent set incidence matrix of $H$; thus $a_{ij}=1$ if $v_i \in I_j$, and $a_{ij}=0$ otherwise.  The fractional chromatic number of the weighted hypergraph $(H, \tau)$, denoted by $\chi_f(H,\tau)$, is the value of the linear program: min $1^T t$ subject to $Mt \ge \tau$, $t \ge 0$.  It can be seen that a demand vector $\tau$ is feasible if and only if $\chi_f(H, \tau) \le 1$.  Next, we shall give a  sufficient condition for  admission control that can be implemented in a distributed manner. 

Let $H=(V,\mathcal{E})$ be a hypergraph. For $i \in V$, define the set $N(i)$ of neighbors of vertex $i$ by
$$N(i) := \{j \in V: \{i,j\} \subseteq E \mbox{ for some } E \in \mathcal{E} \}.$$
For $i \in V$, define

$$\Delta_{ij} = \begin{cases}
\mbox{max} \left\{ \frac{1}{|E|-1}: E \in \mathcal{E}, \{i,j\} \subseteq E \right\}, &\mbox{if } j \in N(i) \\
0, &\mbox{if } j \notin N(i)
\end{cases}$$

\begin{Theorem}  \label{thm:LiNegi:suff:condition}
\cite{Li:Negi:2012} Let $H$ be a hypergraph, with $\Delta_{ij} = \Delta_{ij}(H)$ as defined above.  A sufficient condition for a demand vector $\tau$ to be feasible is that
$$\tau(v_i) + \sum_{j \ne i} \left\{ \Delta_{ij} \tau(v_j) \right\} \le 1, \forall i \in [N].$$
\end{Theorem}

Theorem~\ref{thm:LiNegi:suff:condition} gives a distributed mechanism for admission control and yields a so-called maximal scheduling algorithm.  The actual resource requirement for satisfying demand $\tau$ is given by $\chi_f(H, \tau)$. An upper bound estimate for this resource requirement, as given by Theorem~\ref{thm:LiNegi:suff:condition}, is 
$$B(H, \tau) = \max_{i \in [N]} \left\{ \tau(v_i) + \sum_{j \ne i } \left\{ \Delta_{ij} \tau(v_j) \right\} \right\}.$$
We know $\chi_f(H, \tau) \le B(H, \tau)$.  The distributed maximal scheduling algorithm is suboptimal because it can overestimate the resource requirements needed to satisfy a given demand vector in the sense that it is possible that $\chi_f(H, \tau) \le 1$, but $B(H, \tau) > 1$. In the worst case, this sufficient condition for feasibility overestimates the resource requirements by a factor equal to $\eta(H)$, where
$$\eta(H) = \sup_{\tau \ne 0} \frac{B(H, \tau)}{\chi_f(H, \tau)}.$$
It turns out this worst case performance is exactly equal to the interference degree of the hypergraph, defined next.

\begin{Definition} \label{def:interference:degree:hyp} \cite{Ganesan:TIT:2021} 
The interference degree $\sigma(H)$ of a hypergraph $H=(V,\mathcal{E})$ is defined as follows.  
For $i \in V$, define
$$\Delta_i' := \max_{J \subseteq N(i): J \in \mathcal{I}(H)} \sum_{j \in J} \Delta_{ij},$$
and define
$$\Delta_i'' := \max_{J \subseteq N(i): J \cup \{i\} \in \mathcal{I}(H)} 1 + \sum_{j \in J}\Delta_{ij}.$$
Let $\Delta' := \max_{i \in V} \Delta_i'$, and $\Delta'':=\max_{i \in V} \Delta_i''$.  Then, the interference degree of the hypergraph $H$ is defined by
$$\sigma(H) := \max \{ \Delta', \Delta'' \}.$$
\end{Definition}

The following theorem highlights the importance of the interference degree $\sigma(H)$:

\begin{Theorem} 
\label{thm:beta:equals:sigma} \cite{Ganesan:TIT:2021}
Let $H=(V, \mathcal{E})$ be a conflict hypergraph. The worst-case performance $\eta(H)$ of the distributed maximal scheduling algorithm is equal to the interference degree $\sigma(H)$. 
\end{Theorem}

\subsection{Notation and Terminology} \label{sec:notation}

The notation and terminology on hypergraphs used in the sequel is standard; any further terminology not explicitly recalled here can be found in \cite{Bollobas:1986} \cite{Berge:1985} \cite{Berge:1989}.  In the sequel, $[N]$ denotes the set $\{1,2,\ldots,N\}$.

A hypergraph is said to be \emph{realizable} if it arises as the interference model of a wireless network, i.e. if it  is generated by some wireless network in the sense of Definition~\ref{def:hypergraph:generated:by:wireless}.  

Let $H=(V, \mathcal{E})$ and $H'=(V',\mathcal{E}')$ be hypergraphs.  The hypergraphs $H$ and $H'$ are \emph{isomorphic} if there exists a bijection $f$ from $V$ to $V'$ such that $f(\mathcal{E}) = \mathcal{E}'$, where $f(\mathcal{E}) := \{f(E): E \in \mathcal{E} \}$ and $f(E)$ $(E \subseteq V)$ is the image $\{f(x): x \in E\}$ of $E$ under $f$.  Thus, two hypergraphs are isomorphic if they are the same hypergraph up to a relabeling of the points.

If $S$ is a finite set, then $S^{(r)}$ denotes the collection of all $r$-subsets of $S$.  A hypergraph $(S,\mathcal{E})$ is said to be \emph{$r$-uniform} if the size of each hyperedge is $r$, i.e. if $\mathcal{E} \subseteq S^{(r)}$.   Thus, a $2$-uniform hypergraph is just a graph.   The \emph{complete $r$-uniform hypergraph of order $n$}, denoted by $K_n^{(r)}$, is defined to be the hypergraph with vertex set $[n]$ and with the edge set equal to the collection $[n]^{(r)}$ of all $r$-subsets of $[n]$.


\section{Related Work} \label{sec:related:work}

Many researchers have investigated the structure of conflict graphs of wireless networks under additional assumptions  that are based on geometric, physical (SINR), or graph-theoretic considerations, and exploited this structure for algorithm design.   

For example, the unit disk graph assumption has been well-studied by researchers in the sensor networks community.  In \cite{MaratheEtAl:1995}, an algorithm is given that has a performance guarantee of $3$ for the maximum independent set problem for unit disk graphs, even though this problem is hard to approximate for general graphs. In \cite{MaratheEtAl:1995}, an online algorithm for vertex coloring is given that has a competitive ratio of $6$ for unit disk graphs.  The interference degree of a unit disk graph is at most $5$, and so the maximal scheduling algorithm is a factor of at most $5$ away from optimal for unit disk graphs \cite{Ganesan:2009} \cite{Ganesan:WN:2014}.  A distributed algorithm for admission control known as the scaled clique constraints is a factor of at most $2.2$ away from optimal for unit disk graphs \cite{Gupta:Musacchio:Walrand:07}.  

Under the primary interference model (also known as the $1$-hop interference model or the node-exclusive interference model), the conflict graph is a so-called line graph \cite{Godsil:Royle:2001}.  In this case, the optimal, centralized version of the admission control and scheduling problems can be solved in polynomial time \cite{Hajek:Sasaki:1988}. Under the primary interference model, distributed algorithms that rely only on localized information have been proposed and their performance has been analyzed \cite{Shannon:1949} \cite{Kodialam:Nandagopal:05} \cite{Ganesan:2009} \cite{Ganesan:WN:2014} \cite{Ganesan:ToN:2020} \cite{Ganesan:WiSPNET:2020}.  Secondary interference models are investigated in \cite{Balakrishnan:etal:2004} \cite{Ganesan:ICDCN:2021}, and the $K$-hop interference model is investigated in \cite{Sharma:Mazumdar:Shroff:2006}.   

Given any conflict graph, the worst case performance of the distributed admission control and scheduling algorithm known as maximal scheduling was analyzed in \cite{Chaporkar:Kar:Luo:Sarkar:2008} and independently in \cite{Ganesan:2008} \cite{Ganesan:2009} \cite{Ganesan:2010} and shown to be equal to the interference degree of the conflict graph.  The worst-case performance of the maximal scheduling algorithm under the hypergraph interference model was analyzed in \cite{Ganesan:TIT:2021} and was shown to be equal to the interference degree of the hypergraph.  

Under a certain unit disk graph-based model called the bidirectional (equal power) communication model, the interference degree of the conflict graph is at most $8$ \cite{Chaporkar:Kar:Luo:Sarkar:2008}.   Line networks, wherein all nodes lie on the same line, were studied in \cite{Kose:Medard:PIMRC:2017}.  Under the protocol interference model (cf. \cite{Gupta:Kumar:2000}), the interference degree of the conflict graph of certain line networks was shown to be at most $3$, and this bound is best possible \cite{Ganesan:ToN:2020}.  A \emph{claw} in a graph is an induced subgraph isomorphic to $K_{1,3}$.   When the interference degree of a graph is at most $2$, the graph is claw-free; in this case, efficient algorithms exist for many problems, including the maximum independent set problem \cite{Minty:1980} \cite{Nakamura:Tamura:2001} \cite{Faenza:etal:2014}.  Recently, it was shown that the interference degree of conflict graphs arising in NOMA cellular networks is at most $2$ \cite{Kose:etal:IEEECommLett:2021}.  

In the protocol interference model of \cite{Gupta:Kumar:2000}, the transmission range of a node is a unit ball centered at that node's location.  In \cite{Avin:Emek:etal:JACM:2012}, it is shown that the unit disk graph model can sometimes be too optimistic (cf. \cite[Fig. 2]{Avin:Emek:etal:JACM:2012}) and sometimes too pessimistic (cf. \cite[Fig. 4]{Avin:Emek:etal:JACM:2012}), and SINR diagrams that partition the plane into reception zones are investigated.  The authors show that when all stations are simultaneously active, the reception zones satisfy certain properties; this is a different problem from the one studied in \cite{McEliece:Sivarajan:1994} \cite{Sarkar:Sivarajan:1998} and in the present work, where only certain subsets of stations can be simultaneously active.
The $2$-hop interference model was studied in \cite{Ganesan:TCS:2023}.

Hypergraph models for cellular systems were investigated in \cite{McEliece:Sivarajan:1991} \cite{McEliece:Sivarajan:1994} \cite{Sarkar:Sivarajan:1998} \cite{Ganesan:ICDCN:23}, which discuss the physical (or SINR) model considered in the present work.

\section{The Interference Degree of Structured Hypergraphs} \label{sec:interference:degree:hyp}

In the present section we investigate certain properties of a hypergraph invariant - the interference degree.  We show that the problem of computing the interference degree of a hypergraph is NP-hard, and we prove some basic properties of this hypergraph invariant and compute the interference degree of certain structured hypergraphs. These results extend or generalize the previous results in the literature on graph models to hypergraph models.

\begin{Theorem} \label{thm:graph:interference:degree:NPhard} 
The problem of computing the interference degree of a graph is NP-hard.
\end{Theorem}

\noindent \emph{Proof:}
The proof is by reduction from the independence number problem.  The independence number $\alpha(G)$ of a graph $G$ is the maximum number of vertices that are pairwise nonadjacent. The problem of computing the independence number  of a graph is NP-hard \cite{Garey:Johnson:1979}; in fact, even approximating the independence number to within a constant factor is computationally hard.  Given a graph $G=(V,E)$, construct an augmented graph $G'$, which is obtained from $G$ by joining a new vertex $x$ to each vertex of $G$.  Thus, $G'$ has $|V|+1$ vertices and $|E|+|V|$ edges.  The interference degree $\sigma(G')$ of the augmented graph $G'$ is equal to the independence number $\alpha(G)$. Hence, $\alpha(G) \ge k$ if and only if $\sigma(G') \ge k$.  
\qed

\begin{Theorem} 
The problem of computing the interference degree of a hypergraph is NP-hard.
\end{Theorem}

\noindent \emph{Proof:}
Given any graph $G=(V,E)$, construct a $2$-uniform hypergraph $H=(V,E)$ with the same set of edges as $G$.  Then, the interference degree of the hypergraph $H$ is equal to the interference degree of the graph $G$ \cite[p. 2955]{Ganesan:TIT:2021}. Hence, the problem of computing the interference degree of a graph is a special case of the problem of computing the interference degree of a hypergraph.  The assertion follows from Theorem~\ref{thm:graph:interference:degree:NPhard}. 
\qed

While the problem of computing the interference degree of a graph is NP-hard in general, the problem of computing the interference degree of a unit disk graph can be solved in polynomial time:  given a unit disk graph $G$, since its interference degree is at most $5$, it suffices to examine, for each vertex $v \in V(G)$, the $5$-subsets in its set $\Gamma(v)$ of neighbors, and compute the maximum size of an independent set in each of these $5$-subsets.

In the study of interference degree of graphs, two extreme examples that arise are the star graph $K_{1,r}$, which has the largest possible interference degree for a given number of vertices, and the complete graph $K_n$, which has the smallest possible interference degree for any graph.  The result $\sigma(K_{1,r})=r$ has been generalized to hypergraphs in \cite[Corollary 9]{Ganesan:TIT:2021}, where the interference degree of $\beta$-star hypergraphs was determined.  The next result shows that the interference degree of the complete $r$-uniform hypergraph $K_n^{(r)}$ is $2 - \frac{1}{r-1}$; the special case of this result when $r=2$  is the fact (cf. \cite[p. 303]{Ganesan:2009} \cite[p.1326]{Ganesan:WN:2014}) that the interference degree of the complete graph $K_n$ is $1$. 

\begin{Lemma} 
The interference degree of the complete $r$-uniform hypergraph $H=K_n^{(r)}$ $(2 \le r \le n)$ is 
$$\sigma(H) = 2 - \frac{1}{r-1}.$$
\end{Lemma}

\noindent \emph{Proof:}
Consider the complete $r$-uniform hypergraph $K_n^{(r)}$, which has vertex set $[n]$ and edge set $[n]^{(r)}$.  
By Definition~\ref{def:interference:degree:hyp}, we need to determine $\Delta_i'$ and $\Delta_i''$ ($i \in [n]$).  By symmetry, these values are independent of $i$. So fix $i \in [n]$.  Since the hypergraph is complete, each vertex is in the neighborhood set of all other vertices, i.e. $N(i) = V - \{i\}$.  Each hyperedge has size $r$, whence $\Delta_{ij} = \frac{1}{r-1}$, for all $j \ne i$.  The hyperedges are exactly the $r$-subsets of the vertex set, which implies that the maximum size of an independent set $J$ in $N(i)$ is $r-1$.  Hence, 
$$\Delta_i' = \max_{J \subseteq N(i): J \in \mathcal{I}(H)} \sum_{j \in J} \Delta_{ij}  = (r-1) \frac{1}{r-1} = 1.$$
The maximum size of an independent set $J$ in $N(i)$ such that $J \cup \{i\}$ is an independent set is $r-2$. Hence, 
$$\Delta_i'' = \max_{J \subseteq N(i): J \cup \{i\} \in \mathcal{I}(H)} 1 + \sum_{j \in J} \Delta_{ij} = 1 + \frac{r-2}{r-1} = 2 - \frac{1}{r-1}.$$
By Definition~\ref{def:interference:degree:hyp},
$$\sigma(H) = \max\{1, 2-\frac{1}{r-1} \} = 2 - \frac{1}{r-1}.$$
\qed

In the special case  $r=n$, we get that the interference degree of the hypergraph $([n], \{[n] \})$ consisting of a single hyperedge is $2-\frac{1}{n-1}$.  This gives the interference degree of hypergraphs $H=(V,\mathcal{E})$ when $|\mathcal{E}|=1$.

The interference degree of certain structured hypergraphs such as $\beta$-stars and complete uniform hypergraphs have been obtained. The question of whether these structured hypergraphs are realizable is a future research direction.

The independence number of a hypergraph $H$, denoted by $\alpha(H)$, is the maximum size of an independent set in $H$.  It's clear from Definition~\ref{def:interference:degree:hyp} that $\sigma(H) \le \alpha(H)$:

\begin{Lemma} \label{lem:sigma:le:alpha} 
The interference degree of a hypergraph is at most its independence number.
\end{Lemma}

\noindent \emph{Proof:}
Let $H=(V,\mathcal{E})$ be a hypergraph.  By Definition~\ref{def:interference:degree:hyp}, $\Delta_{ij} \le 1$, for all $i, j \in V$.  Hence, for $i \in V$, 
$$|\Delta_i'| \le \max_{J \subseteq N(i): J \in \mathcal{I}(H)} |J| \le \alpha(H),$$
and 
$$|\Delta_i''| \le \max_{J \subseteq N(i): J \cup \{i\} \in \mathcal{I}(H)} 1 + |J| \le \alpha(H).$$
Hence, $\sigma(H) \le \alpha(H)$.
\qed

The minimum and maximum possible values of a hypergraph with ground set $V=[n]$ is obtained next.

\begin{Lemma} 
Let $H=(V,\mathcal{E})$ be a hypergraph with ground set $V=[n]$.  Then, 
$$1 \le \sigma(H) \le n-1,$$
and both bounds are tight.
\end{Lemma}

\noindent \emph{Proof:}
By definition of $\Delta_i''$, we have $\sigma(H) \ge 1$. If the edge set $\mathcal{E}$ is empty, then the neighborhood sets $N(i)$ are empty, and so $\Delta_i''=1$ for all $i$ and $\sigma(H)=1$. If the edge set is nonempty, then the maximum size of an independent set in $H$ is at most $n-1$.  By Lemma~\ref{lem:sigma:le:alpha}, $\sigma(H) \le n-1$.  The star hypergraph $K_{1,n-1}$ attains the upper bound, and the complete $2$-uniform hypergraph $K_n^{(2)}$ attains the lower bound.
\qed


\section{Non-Realizable Hypergraphs: Preliminaries} \label{sec:preliminaries}

In the special case the hypergraph $(V,\mathcal{E})$ is $2$-uniform, its interference degree is equal to the interference degree of the graph $(V,\mathcal{E})$ \cite[p. 2955]{Ganesan:TIT:2021}.  Thus, the interference degree of the $2$-uniform hypergraph $K_{1,r}$ is $r$.  In light of these results, when investigating performance guarantees of the distributed maximal scheduling algorithm for hypergraph interference models, a question that arises naturally is: what is the maximal value of $r$ such that the hypergraph $K_{1,r}$ is realizable?  This question is investigated in subsequent sections for various values of the path loss exponent $\gamma$.  In this section, we first show that hypergraph models can be viewed as a generalization of the unit disk graph model (cf. Lemma~\ref{lem:udg:special:case:of:hypergraph}).  We will then prove some preliminary results that hold for any value of the path loss exponent $\gamma$ and that will be used in the proofs in subsequent sections.  In the rest of this section, the path loss exponent $\gamma$ can be assumed to be any value greater than or equal to $1$. 

Definition~\ref{def:hypergraph:generated:by:wireless} of the hypergraph of a wireless network $\mathcal{A} = \langle S, \gamma, \beta \rangle$ can be generalized by restricting the maximum size of a hyperedge to be at most some value $k$.  In the special case $k=|S|$, there is no restriction on the size of a hyperedge, and so we get Definition~\ref{def:hypergraph:generated:by:wireless} above.  In the special case $k=2$, the resulting hypergraph is a unit disk graph, and conversely (see Lemma~\ref{lem:udg:special:case:of:hypergraph} below) every unit disk graph can be obtained as the hypergraph of a wireless network with $k=2$; thus, we can view the unit disk graph construction as a special case of a more general method of construction.  

\begin{Lemma} \label{lem:udg:special:case:of:hypergraph}
Suppose $\mathcal{A} = \langle S, \gamma, \beta \rangle$ is a wireless network, where $\gamma > 0$ and $\beta=1$.  Let $G = (S,E)$ be the unit disk graph generated by $S$, and let $H = (S, \mathcal{E})$ be the hypergraph generated by $\mathcal{A}$.  Then, $\mathcal{E} \cap S^{(2)} = E$. 
\end{Lemma}

\noindent \emph{Proof:}
Let $\{x,y\} \subseteq S$.  The Euclidean distance $d(x,y) \le 1$ if and only if $\frac{1}{d(x,y)^\gamma} \ge 1$, which is the case if and only if $\{x,y\}$ is a forbidden set.  Every $2$-subset of $S$ that is forbidden is a minimal forbidden set and belongs to $\mathcal{E}$.  Thus, $\{x,y\} \in E$ if and only if $\{x,y\} \in \mathcal{E}$.  Thus, $E = \mathcal{E} \cap S^{(2)}$ and $E \subseteq \mathcal{E}$.
\qed

As a corollary, we obtain the following: Given any unit disk graph $G=(S,E)$, there exists a wireless network $\mathcal{A} = \langle S, \gamma, \beta \rangle$ such that the hypergraph $H=(S,\mathcal{E})$ generated by $\mathcal{A}$ satisfies $\mathcal{E} \cap S^{(2)} = E$.  Conversely, given any wireless network $\mathcal{A} = \langle S, \gamma, \beta \rangle$ and the hypergraph $H=(S,\mathcal{E})$ generated by $\mathcal{A}$, the graph $G=(S, \mathcal{E} \cap S^{(2)})$ is a unit disk graph.

Let $\mathcal{A} = \langle S, \gamma, \beta \rangle$ be a wireless network and let $\pi: \mathbb{R}^2  \rightarrow \mathbb{R}^2, s \mapsto s'$ be any map that preserves distances.  For example, $\pi$ could be a rigid motion such as  a rotation, a reflection, or a translation. Let $S' := \{\pi(s): s \in S\}$ be the set of new locations of the stations.  Because $\pi$ preserves distances, it is clear that the hypergraph generated by $\mathcal{A} = \langle S, \gamma, \beta \rangle$ and the hypergraph generated by $\mathcal{A}' = \langle S', \gamma, \beta \rangle$ are isomorphic.  

\begin{Lemma} \label{lem:isomorphic:scaling}   
Given a wireless network $\mathcal{A} = \langle S, \gamma, \beta \rangle$ and a positive real number $\rho$, define the scaled wireless network $\mathcal{A}' := \langle S', \gamma, \beta' \rangle$, where $S'= \{s': s \in S\}$, $s'=T_\rho(s)$ is the new location of station $s$ obtained using the scaling transformation $T_\rho: \mathbb{R}^2 \rightarrow \mathbb{R}^2$ that maps $s=(x,y)$ to $s':= (\rho x, \rho y)$, and $\beta' := \frac{\beta}{\rho^\gamma}$.  Then, the hypergraphs generated by the wireless networks $\mathcal{A}$ and $\mathcal{A}'$ are isomorphic.
\end{Lemma}

\noindent \emph{Proof:} 
Let $W \subseteq S$, and let $W' \subseteq S'$ be the image of $S$ under the scaling transformation $T_\rho$.  It suffices to show that $W$ is a forbidden set for the wireless network $\mathcal{A}$ if and only if $W'$ is a forbidden set for the wireless network $\mathcal{A}'$, because that would imply the  map $f: S \rightarrow S', s \mapsto s'$ is an isomorphism between the corresponding hypergraphs.  

Fix $w \in W$.  Then, the energy at $w'$ due to transmissions by the set of stations $W' - \{w'\}$ is 
\begin{align*}
E(W' - \{w'\}, w') & = \sum_{s' \in W' - \{w'\}} \frac{1}{d(s',w')^\gamma} \\
& = \sum_{s \in W - \{w\}} \frac{1}{(\rho d(s,w))^\gamma} \\
& = \frac{1}{\rho^\gamma} E(W-\{w\},w).
\end{align*}
Thus, $W'$ is a forbidden subset for the wireless network $\mathcal{A}'$ if and only if $W$ is a forbidden subset for the wireless network $\mathcal{A}$.  The map from $S$ to $S'$ that takes $s$ to $s'$ is an isomorphism from the hypergraph generated by $\mathcal{A}$ to the hypergraph generated by $\mathcal{A}'$. 
\qed

In the rest of this section, we prove some results that are used in the proofs in subsequent sections.  The reader may skip the rest of this section now and return to it later.  

The \emph{unit ball} refers to the set $\{(x,y): x^2 + y^2 \le 1\}$ of all points in $\mathbb{R}^2$ that are within unit distance from the origin, and the \emph{unit circle} refers to the boundary $\{(x,y): x^2 + y^2=1\}$ of the unit ball. The unit ball (minus the origin) is the set $\{(x,y): x^2+y^2 \le 1, (x,y) \ne (0,0)\}$.

\begin{Lemma} \label{lem:diameter:60deg:sector} 
The maximum distance between any two points in a $60^\circ$ sector of the unit ball is $1$.
\end{Lemma}

\noindent \emph{Proof:} 
Let $a, b$ be two points in a $60^\circ$ sector of the unit ball, with polar coordinates $a=(r_1,\theta_1)$ and $b=(r_2,\theta_2)$, where $0 \le r_1, r_2 \le 1$, and where $0 \le \theta_1, \theta_2 \le \pi/3$ without loss of generality.  It will be shown that the Euclidean distance $d(a,b)$ is at most $1$.

If $r_1=0$ or $r_2=0$, then $d(a,b) \le 1$ because the radius of the unit ball is $1$. So suppose  $r_1,r_2 > 0$.  Let $\psi = |\theta_1 - \theta_2|$.  By the law of cosines,
\begin{align*}
d(a,b) & = r_1^2 + r_2^2 - 2 r_1 r_2 \cos \psi
\\ 
& \le r_1^2 + r_2^2 - 2 r_1 r_2 \cos \pi/3
\\
& = r_1^2 + r_2^2 - r_1 r_2
\\
& =\frac{r_1^3 + r_2^3}{r_1 + r_2}
\\
& \le 1,
\end{align*} 
where the first inequality holds because $\psi \le \pi/3$, and the second inequality holds because $r_1^3 \le r_1$ and $r_2^3 \le r_2$ when $0 \le r_1, r_2 \le 1$.  Equality holds throughout if and only if $r_1=1, r_2=1$ and $\psi=\pi/3$. 
\qed

We show next that if any two points in the unit ball (minus the origin) are at distance at least $1$, then pushing the two points radially outward to the boundary of the unit ball will not decrease the Euclidean distance between the points.

\begin{Lemma} \label{lem:radially:outward}
Suppose $a$ and $b$ are two points in the unit ball (minus the origin) such that the distance $d(a,b)$ between them is greater than $1$.  Let $a'$ and $b'$ be obtained by moving the points $a$ and $b$, respectively, radially outward to the boundary of the unit ball.  Thus, if the polar coordinates of $a$ and $b$ are $(r_1, \theta_1)$ and $(r_2, \theta_2)$, respectively, where $0 < r_1, r_2 \le 1$, then the polar coordinates of $a'$ and $b'$ are   $(1, \theta_1)$ and $(1, \theta_2)$, respectively.  
Then, $d(a', b') \ge d(a, b)$.
\end{Lemma}

\noindent \emph{Proof:} Let $a, b$ be points with polar coordinates $a=(r_1,\theta_1)$, $b=(r_2,\theta_2)$, where $0 < r_1, r_2 \le 1$ and such that $d(a,b) \ge 1$.  Define the points $a':=(1,\theta_1)$ and $b' :=(1, \theta_2)$.  It suffices to prove that $d(a',b') \ge d(a,b)$.

If $r_1,r_2 < 1$, then define $\rho := \frac{1}{\max\{r_1,r_2\}}$.  Because $\rho > 1$, the transformation $T_\rho$ that maps the point $(r,\theta)$ to $(\rho r, \theta)$ increases the distance between any two points, i.e. $T_\rho$ satisfies the property that for any two points $a$ and $b$,  $d(T_\rho(a), T_\rho(b)) \ge d(a,b)$.  Thus, it can be assumed without loss of generality that at least one of the two points $a$ or $b$, $a$ say, lies on the boundary of the unit ball.  Thus, $r_1=1$. 

Because a rotation of the Euclidean plane preserves distances, it can be assumed that $\theta_1=0$.  So, in polar form, the coordinates of $a$ and $b$ can be assumed to be $a=(1,0)$ and $b=(r_2,\theta_2)$, where $0 < r_2 \le 1$, and $\theta_2 > \pi/3$ because $d(a,b) > 1$.  By symmetry about the horizontal axis, we may assume the point $b$ lies in the first or second quadrant. 

Observe that $a'=(1,0)$ and $b'=(1,\theta_2)$.  Define $D :=d(a,b)$ and $D':=d(a',b')$.  Then, $D = \sqrt{(r_2 \cos \theta_2 -1)^2 + (r_2 \sin \theta_2)^2}$ and $D' = \sqrt{(\cos \theta_2-1)^2+(\sin \theta_2)^2}$.  It needs to be proved that $D' \ge D$.

We have that 
\begin{align*}
(D')^2 - D^2 & = 1 - 2 (1-r_2) \cos \theta_2 - r_2^2
\\
& \ge 1 - (1-r_2)-r_2^2 
\\
& \ge 0,
\end{align*}
where the first inequality holds because $\theta > \pi/3$, and the second inequality holds because $r_2 \le 1$.  
\qed

Next, we show that in determining which hypergraphs are realizable, it can be assumed without loss of generality that the reception threshold $\beta$ is equal to $1$.

\begin{Lemma} \label{lem:wlog:beta:equals:1}    
Let $H$ be a hypergraph. If $H$ is realizable, then $H$ is realizable for some wireless network $\mathcal{A} = \langle S, \gamma, \beta \rangle$ having $\beta = 1$. 
\end{Lemma}

\noindent \emph{Proof:}
Suppose the hypergraph $H$ is realizable.  Then, there exists a wireless network $\mathcal{A} = \langle S, \gamma, \beta \rangle$ that generates $H$.  Define the scaling factor $\rho := e^{\frac{\ln \beta}{\gamma}}$.  Let $\mathcal{A}' = \langle S', \gamma, \beta' \rangle$ be the wireless network obtained by scaling the wireless network $\mathcal{A}$ by a factor $\rho$ in the manner defined in the statement of Lemma~\ref{lem:isomorphic:scaling}.   By Lemma~\ref{lem:isomorphic:scaling}, the wireless network $\mathcal{A}' = \langle S', \gamma, 1 \rangle$ generates $H$. 
\qed

\section{Nonrealizability of the Hypergraph $K_{1,5}$ When $\gamma=4$}
\label{sec:alpha:4}

The structure of hypergraphs generated by wireless networks $\langle S, \gamma, \beta \rangle$ depends on the value of the path loss exponent $\gamma$.  In typical wireless environments, the path loss exponent $\gamma$ takes values in the range from about $1.6$ to $6.5$ \cite{Goldsmith:2020}.  The special case $\gamma=4$ has been studied by various authors \cite{McEliece:Sivarajan:1994}  \cite{Sarkar:Sivarajan:VTC:2002}. In the present section, we recall results from our previous work \cite{Ganesan:ICDCN:23}  on  the structure of realizable and nonrealizable hypergraphs under the signal propagation model $\gamma=4$.  We first show that the $2$-uniform hypergraph $K_{1,4}$ is realizable.  Then, we prove that the $2$-uniform hypergraph $K_{1,5}$ is not realizable.  In other words, it can be shown that for any placement of six stations $S = \{s_0,s_1,\ldots,s_5\} \subseteq \mathbb{R}^2$ in the Euclidean plane and for any reception threshold $\beta$, the hypergraph generated by the wireless network $\langle S, \gamma=4, \beta \rangle$ is not isomorphic to $K_{1,5}$.    Thus, while the maximal value of $r$ such that the graph $K_{1,r}$ is a unit disk graph is $r=5$,  the maximal value of $r$ such that the  hypergraph $K_{1,r}$ is realizable is $r=4$.

\begin{Lemma} \label{lem:K1:4:alpha:4} 
If the path loss exponent $\gamma=4$, then the hypergraph $K_{1,4}$ is realizable.
\end{Lemma}

\noindent \emph{Proof:}
Consider the wireless network $\mathcal{A} = \langle S, \gamma, \beta \rangle$, where $S = \{s_0, s_1, s_2, s_3, s_4\}$, $s_0=(0,0)$, $s_1=(1,0)$, $s_2=(0,1)$, $s_3=(-1,0)$, and  $s_4=(0,-1)$.  Suppose $\gamma=4$ and $\beta=1$.  Then,   $\frac{1}{d(s_0,s_i)^4} \ge \beta$ $(i \in [4])$, and so $\{s_0,s_i\}$ is a forbidden set for all $i \in [4]$.  

We show that $W = \{s_1,s_2,s_3,s_4\}$ is feasible.  By symmetry, it suffices to show that the energy $E(W-\{s_1\}, s_1\}$ at $s_1$ due to transmissions by $\{s_2,s_3,s_4\}$ is less than the reception threshold $\beta$.   We have that
\begin{align*}
E(W-\{s_1\}, s_1) & = \sum_{w \in \{s_2,s_3,s_4\}} \frac{1}{d(w,s_1)^4} \\
& = \frac{1}{(\sqrt{2})^4} + \frac{1}{2^4} + \frac{1}{(\sqrt{2})^4} < 1.
\end{align*}
Since $W$ is feasible, the minimal forbidden sets of the wireless network $\mathcal{A}$ are precisely the subsets $\{s_0,s_i\}$ ($i \in [4]$).  Thus, the hypergraph generated by the wireless network $\mathcal{A}$ is the $2$-uniform hypergraph $K_{1,4}$.
\qed

\begin{Theorem} \label{thm:K1:5:alpha:4} 
If the path loss exponent $\gamma=4$, then the hypergraph $K_{1,5}$ is not realizable.
\end{Theorem}

\noindent \emph{Proof:}
By way of contradiction, suppose there exists a wireless network $\mathcal{A} = \langle S, \gamma, \beta \rangle$ that generates the hypergraph $(S, \mathcal{E}) = K_{1,5}$, where $S = \{s_0, s_1, \ldots, s_5\}$ and $\mathcal{E}$ is the collection $\{ \{s_0,s_i\}: i \in [5] \}$ of five hyperedges. Recall that $\gamma=4$. By Lemma~\ref{lem:wlog:beta:equals:1}, it can be assumed that  $\beta=1$. Since a translation of the Euclidean plane is a rigid motion and preserves distances, it can be assumed that the station $s_0$ is located at the origin.  

Let $W:=\{s_1,s_2,s_3,s_4,s_5\}$.   
Because $\{s_0,s_i\}$ $(i \in [5])$ is a forbidden set, the distance between the stations $s_0$ and $s_i$ is at most $1$. Thus, the set $W$ of stations lies inside the unit ball (minus the origin).  Observe that $W$ does not contain a hyperedge, and hence $W$ is an independent set of the hypergraph.  In other words, for all $w \in W$, the interference is tolerable at $w$ in the sense that 
$$E(W-\{w\},w) = \sum_{s \in W - \{w\}} \frac{1}{d(s,w)^4} < 1.$$  
We will arrive at a contradiction by proving that this inequality is violated for every set of five locations in the unit ball (minus the origin) for $W$. 


For $i, j \in [5], i \ne j$, we have that $\{s_i, s_j\}$ is not a forbidden subset, whence the distance between $s_i$ and $s_j$ is greater than $1$.  Thus, the distance between any two stations in $W$ is greater than $1$.  


The points in $W$ can be pushed radially outward to the boundary of the unit ball, i.e. to the unit circle, and by Lemma~\ref{lem:radially:outward} this transformation does not decrease the distance between any two points in $W$ and hence  preserves feasibility of $W$.  Thus, we can assume that the hypergraph $K_{1,5}$ is generated by a wireless network $\langle S, \gamma, \beta \rangle$, such that $S = \{s_0,s_1,\ldots,s_5\}$, $\gamma=4$, $\beta=1$, and the set of points $W = \{s_1,\ldots,s_5\}$ lies \emph{on} the unit circle.  

\begin{figure}
\centering
\begin{tikzpicture}[scale=3]
  \filldraw[fill=blue!20,draw=red] (0,0) -- (3mm,0mm)
    arc [start angle=0, end angle=70, radius=3mm] -- cycle;

  \filldraw[fill=yellow!20,draw=blue] (0,0) -- (3.5mm,0mm)
    arc [start angle=0, end angle=-70, radius=3.5mm] -- cycle;

    \node[red] at (35:2mm) {$\theta_1$};

    \node[black] at (-35:2mm) {$\theta_5$};

    \draw[->] (-1.2,0) -- (1.2,0) coordinate (x axis)node[right]{$x$};
    \draw[->] (0,-1.2) -- (0,1.2) coordinate (y axis)node[above]{$y$};
    \draw (0,0) circle [radius=1cm];

    \draw (0,0) -- (0.342,0.94); 
    \node at (0.342,0.94)[circle,fill,inner sep=1pt]{};
    \node at (0.35,1) {$s_2$};

    \draw (0,0) -- (0.342,-0.94); 

    \node at (0.39,-1.05) {$s_5$};
    \node at (0.342,-0.94)[circle,fill,inner sep=1pt]{};
    
    \node at (1.1,0.1) {$s_1$};
    \node at (1.0,0)[circle,fill,inner sep=1pt]{};

    \node at (1.5,0.7) {$\theta_1 = \frac{\pi}{3} + \delta_1$};
    
    \node at (1.5,-0.7) {$\theta_5 = \frac{\pi}{3} + \delta_5$};

    \node at (-0.85,0.71) {$s_3$};
    \node at (-0.766, 0.642)[circle,fill,inner sep=1pt]{};

    \node at (-0.85,-0.71) {$s_4$};
    \node at (-0.766, -0.642)[circle,fill,inner sep=1pt]{};
    
\end{tikzpicture}
\caption{Five stations on the unit circle, with stations $s_1,s_2$ and $s_5$ close enough to each other that $\delta_1+\delta_5=2\pi/15$. Relevant to the proof of Theorem~\ref{thm:K1:5:alpha:4}.}
\label{fig:unit:circle:5:points}

\end{figure}
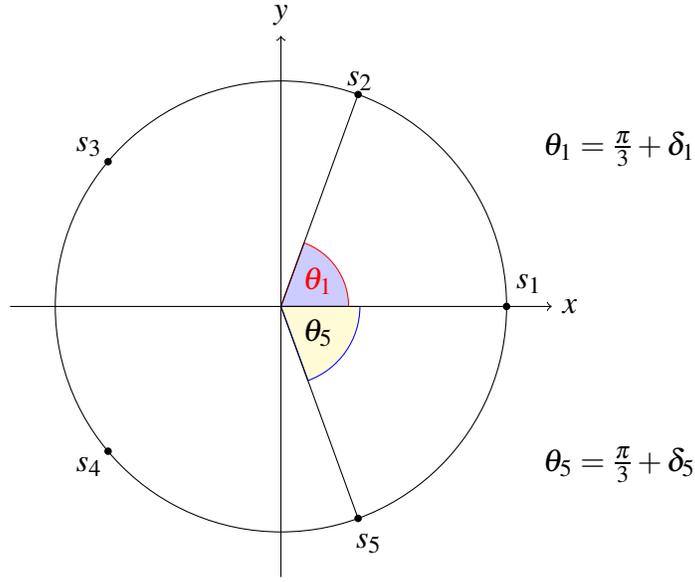

A rotation of the Euclidean plane preserves distances; hence,  we can assume without loss of generality that $s_1$ lies on the unit circle at angle $0$, i.e. that the polar coordinates  of $s_1$ are $(1,0)$.  Without loss of generality, suppose the remaining points $s_2,s_3,s_4,s_5$ lie on the unit circle in clockwise order. Let $\theta_i$ ($i \in [4]$) denote the angle between $s_i$ and $s_{i+1}$, and let $\theta_5$ denote the angle between $s_5$ and $s_1$; see Figure~\ref{fig:unit:circle:5:points}.  Because the distance between any two points in $W$  is greater than $1$, we have that $\theta_i > \pi/3$, for all $i$.  Because the sum of any four of these angles is greater than $4 \pi/3$, each of these angles $\theta_i$ is less than $2 \pi /3$. Thus, $\pi/3 < \theta_i < 2 \pi/3$, for $i \in [5]$.

Define $\delta_i = \theta_i - \pi/3$, for $i \in [5]$.  Because $\sum_{i=1}^5 \theta_i = 2\pi$, we have that $\sum_{i=1}^5 \delta_i = \pi/3$.  Since the sum of these five numbers is $\pi/3$, there must be some two consecutive numbers on the circle, say $\delta_1$ and $\delta_5$, that sum to at most $2 \pi/15$. Thus, there exist three points in $W$, namely $s_1, s_2$ and $s_5$ that are very close to each other in the sense that $\delta_1 + \delta_5 \le 2 \pi/15$.  We will show that the subset $\{s_1, s_2, s_5\}$ is a forbidden subset, even when these three points are farthest apart under the said conditions, i.e. even if $\delta_1 + \delta_5 = 2 \pi/15$.

Thus, we may assume that three of the points of $W$ are at locations  $s_1=(1,0)$, $s_2=(1, \pi/3+\delta_1)$, and $s_5 = (1, -\pi/3 - \delta_5)$, where $\delta_5 = 2 \pi/15 - \delta_1$.  In terms of $\delta_1$,  the interference at $s_1$ due to transmissions by $s_2$ and $s_5$ is given by
$$f(\delta_1) = \frac{1}{d(s_1,s_2)^4} + \frac{1}{d(s_1,s_5)^4},$$
where the domain for $f$ is $[0, 2 \pi/15]$.  By Lemma~\ref{lem:fdelta:greater:than:1}, the minimum value of $f(\delta_1)$  is greater than $1$.  Because the reception threshold $\beta$ is equal to $1$, the subset $\{s_1,s_2,s_5\}$ is a forbidden subset. Hence, its superset $W$ is also a forbidden set, contradicting the fact that $W$ is an independent set.
\qed

\begin{Lemma} \label{lem:fdelta:greater:than:1} 
Let $f(\delta_1)$ be the function
$$
f(\delta_1) := \frac{1}{d(s_1,s_2)^4} + \frac{1}{d(s_1,s_5)^4},
$$
where the points $s_1,s_2$ and $s_5$ have polar coordinates $s_1=(1,0)$, $s_2=(1, \pi/3+\delta_1)$ and $s_5=(1,-\pi/3-\delta_5)$ (as shown in Figure~\ref{fig:unit:circle:5:points}),  $\delta_5 = 2 \pi/15 - \delta_1$, and the domain is $0 \le \delta_1 \le 2 \pi/15$.  
Then, the minimum value of $f(\delta_1)$ is greater than $1$. 
\end{Lemma}

\noindent \emph{Proof:}
Negating the angle of $s_5$ does not change the distance between $s_5$ and $s_1$; so  assume that $s_5 = (1, \pi/3 + \delta_5)$.  Let $\delta := \delta_1 - \pi/15$.  The points $s_1, s_2$ and $s_5$ can be expressed in terms of $\delta$ in polar form as $s_1 = (1,0)$, $s_2 = (1, 2 \pi/5+\delta)$ and $s_5 = (1, 2 \pi/5 - \delta)$.   

Define the function $g(\delta) = d(s_1,s_2)^{-4} + d(s_1,s_5)^{-4}$, where $s_1 = (1,0)$, $s_2 = (1, 2 \pi/5+\delta)$, $s_5 = (1, 2 \pi/5-\delta)$, and the domain is $[-\pi/15, \pi/15]$.  Because $f(\delta_1) = g(\delta + \pi/15)$, it suffices to show that the minimum value of $g(\delta)$  is greater than $1$.

It can be verified that
$$g(\delta) = \frac{1}{4 \left( 1-\cos \left( \frac{2 \pi}{5}+\delta \right) \right)^2} + \frac{1}{4 \left( 1-\cos \left( \frac{2 \pi}{5}-\delta \right) \right)^2}, $$

$$g'(\delta) = - \frac{\sin\left(\frac{2 \pi}{5}+\delta \right)}{2 \left( 1-\cos \left( \frac{2 \pi}{5}+\delta \right) \right)^3} +  \frac{\sin\left(\frac{2 \pi}{5}-\delta \right)}{2 \left( 1-\cos \left( \frac{2 \pi}{5}-\delta \right) \right)^3},  $$
and
$$g''(\delta) = \frac{3-\cos \left(\frac{2\pi}{5}+\delta \right) - 2 \cos^2 \left( \frac{2\pi}{5}+\delta\right)}{2 \left( 1-\cos \left( \frac{2 \pi}{5}+\delta \right) \right)^4} $$ $$+  \frac{3-\cos \left(\frac{2\pi}{5}-\delta \right) - 2 \cos^2 \left( \frac{2\pi}{5}-\delta\right)}{2 \left( 1-\cos \left( \frac{2 \pi}{5}-\delta \right) \right)^4}. $$

Because $3- \cos \gamma  -2\cos^2 \gamma > 0$ when $|\cos \gamma| < 1$,  $g''(\delta) > 0$ for all $\delta$.  Also, $g'(0)=0$.  Thus, $g(\delta)$ is convex and attains a global minimum at $\delta=0$.  It can be verified that  $g(0)=\frac{3+\sqrt{5}}{5}$, which is greater than $1$.
\qed

\begin{Corollary} \label{cor:max:r:K1_r:realizable}
If the path loss exponent $\gamma=4$, then the largest value of $r$ such that the hypergraph $K_{1,r}$ is realizable is $r=4$.
\end{Corollary}

\noindent \emph{Proof:}
The assertion follows from Lemma~\ref{lem:K1:4:alpha:4} and Theorem~\ref{thm:K1:5:alpha:4}.
\qed

Note that Corollary~\ref{cor:max:r:K1_r:realizable} does not imply that the maximum possible interference degree of any realizable hypergraph is $4$.  Determining the realizability of hypergraphs that contain other structures (besides $K_{1,r}$) is a future research direction.

Combining Theorem~\ref{thm:K1:5:alpha:4} and Proposition~\ref{prop:r:nondecreasing}, we obtain a generalization of Theorem~\ref{thm:K1:5:alpha:4} to all nonintegral values of the path loss exponent $\gamma \le 4$: if  the path loss exponent $\gamma \le 4$, then the hypergraph $K_{1,5}$ is not realizable.  

\section{Nonrealizability of the Hypergraph $K_{1,5}$ When $\gamma=3$} \label{sec:alpha:3}

In this section, we consider the case where the path loss exponent $\gamma=3$. It will be shown that the maximal value of $r$ such that the hypergraph $K_{1,r}$ is realizable is $r=4$.

\begin{Lemma} \label{lem:gamma:3:K14:realizable}
If the path loss exponent $\gamma=3$, then the hypergraph $K_{1,4}$ is realizable. 
\end{Lemma}

\noindent \emph{Proof:}
Consider the wireless network $\mathcal{A} = \langle S, \gamma, \beta \rangle$ where $S = \{s_0, s_1, s_2, s_3, s_4\}$, $\gamma=3$, $\beta=1$, $s_0$ is placed at the origin, and the remaining stations $s_i (i=1,\ldots,4)$ are placed uniformly on the unit circle in clockwise order.  Then, $\{s_0, s_i\}$ is a minimal forbidden set for all $i=1,\ldots,4$. The energy at one of the stations in $W = \{s_1,s_2,s_3,s_4\}$, say $s_1$, due to transmissions by the remaining stations in $W$, is given by
\begin{align*}
E(W-\{s_1\},s_1)  & = \frac{1}{d(s_2,s_1)^3} + \frac{1}{d(s_3,s_1)^3}  + \frac{1}{d(s_4,s_1)^3} \\
& = \frac{1}{(\sqrt{2})^3} + \frac{1}{2^3}   + \frac{1}{(\sqrt{2})^3} \\
& = \frac{1}{\sqrt{2}} + \frac{1}{8} \\
& < 1.
\end{align*}
Hence, $W$ is feasible.  It follows that the hypergraph generated by the wireless network $\mathcal{A}$ is the $2$-uniform hypergraph $K_{1,4}$. 
\qed

Intuitively, as the path loss exponent $\gamma$ increases, the signal attenuates more rapidly with distance, leading to reduced interference at the receiver. This makes the realization of $K_{1,5}$ more likely. Conversely, as $\gamma$ decreases, the likelihood of realizing $K_{1,5}$ diminishes. This observation is captured in Proposition~\ref{prop:r:nondecreasing}.  Since $K_{1,4}$ is realizable even at lower values of $\gamma$, such as $\gamma = 3$, it remains realizable at higher values, such as $\gamma = 4$. In other words, Lemma~\ref{lem:gamma:3:K14:realizable} implies Lemma~\ref{lem:K1:4:alpha:4}. Similarly, because $K_{1,5}$ is not realizable even at larger values of $\gamma$, such as $\gamma = 4$, it is also not realizable at $\gamma = 3$:

\begin{Theorem} \label{thm:alpha:3}
If the path loss exponent $\gamma \le 3$, then the hypergraph $K_{1,5}$ is not realizable.  
\end{Theorem}

\noindent \emph{Proof:} The assertion follows from Theorem~\ref{thm:K1:5:alpha:4} and Proposition~\ref{prop:r:nondecreasing}.
\qed

By Theorem~\ref{thm:alpha:3}, when the path loss exponent $\gamma=3$,  the hypergraph $K_{1,5}$ is not realizable.  Combined with Lemma~\ref{lem:gamma:3:K14:realizable}, this leads to the conclusion that for $\gamma=3$, the maximal value of $r$ for which the hypergraph $K_{1,r}$ is realizable is $r=4$.

\section{Nonrealizability of the Hypergraph $K_{1,4}$ When $\gamma \le 2$} \label{sec:alpha:2}

In the present section, we show that when $\gamma=2$, the maximal value of $r$ such that the hypergraph $K_{1,r}$ is realizable is $r=3$. 

\begin{Lemma}
If the path loss exponent $\gamma=2$, then the hypergraph $K_{1,3}$ is realizable.  
\end{Lemma}

\noindent \emph{Proof:}
Let $S=\{s_0, s_1, s_2, s_3\}$ be a set of stations such that $s_0=(0,0)$, and $s_i$ has polar coordinates $(r,\theta_i)=(1, 2 \pi i/ 3)$  $(i=1,2,3)$.  Thus, the latter three points are spaced uniformly on the unit circle. Observe that the distance between $s_0$ and $s_i$ is at most $1$, for all $i$.  Let $\beta=1$ and recall $\gamma=2$. It follows that hypergraph generated by the wireless network $\langle S, \gamma, \beta \rangle$ is $(S,\mathcal{E})$, where $\mathcal{E}$ contains three edges $\{s_0, s_1\}$, $\{s_0, s_2\}$, and $\{s_0, s_3\}$. 

Let $W = \{s_1, s_2, s_3\}$.  Pick any $w \in W$.  It can be verified that the energy $E(W-\{w\}, w\}$ at $w$ due to transmissions by the other two stations is $2/3$, which is less than $1$.  Hence, $W$ is feasible, which implies that there are no other minimal forbidden sets other than the three edges mentioned above. Thus, the hypergraph generated by the wireless network $\langle S, \gamma, \
\beta \rangle$ is $K_{1,3}$. 
\qed

\begin{Theorem} \label{thm:alpha:2}
If the path loss exponent $\gamma  \le 2$, then the hypergraph $K_{1,4}$ is not realizable. 
\end{Theorem}

\noindent \emph{Proof:}
Fix $\gamma \le 2$. By way of contradiction, suppose there exists a set of locations $S=\{s_0, s_1, s_2, s_3, s_4\}$ in the Euclidean plane, with $\beta=1$, such that the wireless network $\langle S, \gamma, \beta \rangle$ generates the hypergraph $K_{1,4}$.  Without loss of generality, we may assume $s_0$ is the center vertex of the star $K_{1,4}$ and we may assume $s_0$ is at the origin.  Thus, $W=\{s_1, s_2, s_3, s_4\}$ is feasible and the points in $W$ lie on the unit ball (minus the origin).  Also, the distance between any two distinct points in $W$ is greater than $1$.  By Lemma~\ref{lem:radially:outward}, the points in $W$ can be pushed radially outward to the boundary of the unit ball (namely the unit circle), and since this only increases the distance between points in $W$, the resulting set of four points on the unit circle is a feasible set.  

Thus, there exists a set $W=\{s_1, s_2, s_3, s_4\}$ of four points on the unit circle such that the distance between any two points in $W$ is greater than $1$ and $W$ is feasible.  It will be shown that this is impossible. 

Without loss of generality, choose a station in $W$, say $s_1$. Because the interference at $s_1$ due to senders in $W-\{s_1\}$ is tolerable, we have that
$$ \sum_{s_j \in W - \{s_1\}} \frac{1}{d(s_j,s_1)^\gamma} < 1.$$
The distance between any two stations on the unit circle is at most $2$,  so each of the three terms in the sum above is at least $\frac{1}{4}$.  By the inequality above, each term in the sum above is less than $\frac{1}{2}$.  In particular, if $s_j$ is the station in $W - \{s_1\}$ that is closest to $s_1$, then $\frac{1}{d(s_j,s_1)^\gamma} < \frac{1}{2}$, i.e. $d(s_j,s_1) >  \sqrt{2}$.  This implies the angle between $s_1$ and its nearest neighbor $s_j$ is more than $90$ degrees. Since $s_1$ was chosen without loss of generality, this conclusion holds for all stations in $W$. In other words, there are four stations on the unit circle such that the angle between any two consecutive stations is more than $90$ degrees, an impossibility.
\qed

\section{Nonrealizability of the Hypergraph $K_{1,3}$ When $\gamma \le 1$} \label{sec:alpha:1}

In this section, we consider the case where the path loss exponent $\gamma \le 1$.  Motivated by results on the performance guarantees of distributed algorithms in wireless networks, we determine the maximal value of $r$ such that the $2$-uniform hypergraph $K_{1,r}$ is realizable.  This value is shown to be equal to $2$.

\begin{Lemma}
If the path loss exponent $\gamma > 0$, then the hypergraph $K_{1,2}$ is realizable.
\end{Lemma}

\noindent \emph{Proof:}
Fix $\gamma > 0$. Consider the wireless network $\mathcal{A} = \langle S, \gamma, \beta \rangle$, where $S = \{s_0, s_1, s_2\}$, $\beta=1$, and in rectangular form the locations of the stations are $s_0 = (0,0)$, $s_1=(1,0)$, and $s_2=(-1,0)$.  Let $H=(S,\mathcal{E})$ be the hypergraph generated by the wireless network $\mathcal{A}$. Because the distance $d(s_0,s_i) \le 1$, $\{s_0,s_i\} \in \mathcal{E}$ ($i=1,2$).  The energy at $s_1$ due to transmission by $s_2$ is given by
$$E(\{s_2\},s_1) = \frac{1}{d(s_1,s_2)^\gamma} = \frac{1}{2^\gamma} < 1.$$
Hence, $\{s_1,s_2\}$ is not a forbidden set. It follows that the hypergraph generated by the wireless network $\mathcal{A}$ is the $2$-uniform hypergraph $K_{1,2}$. 
\qed

\begin{Theorem}
If the path loss exponent $\gamma \le 1$, then the hypergraph $K_{1,3}$ is not realizable.
\end{Theorem}

\noindent \emph{Proof:}
Let $\gamma \le 1$.  By way of contradiction, suppose the hypergraph $K_{1,3}$ is realizable.  Then, using the same arguments as in the proof of Theorem~\ref{thm:alpha:3}, it can be shown that there exists a set $W = \{s_1,s_2,s_3\}$ of three stations on the unit circle such that $W$ is feasible. The distance between any two points in $W$ is at most $2$.  Thus, $d(s_1,s_2) \le 2$ and $d(s_1,s_3) \le 2$.  The interference at $s_1$ due to transmissions by $s_2$ and $s_3$ is 
$$E(\{s_2,s_3\},s_1) = \frac{1}{d(s_1,s_2)^\gamma} + \frac{1}{d(s_1,s_3)^\gamma} \ge \frac{1}{2} + \frac{1}{2} = 1.$$
This implies $W$ is a forbidden set, a contradiction.
\qed

\section{Extensions to Non-integral Values of the Path Loss Exponent $\gamma$} \label{sec:nonintegral:alpha}

In typical wireless scenarios, the path loss exponent is between $1.6$ and $6.5$ \cite{Goldsmith:2020}.  In the present section, we prove some results that hold for all $\gamma > 0$. 

It is well-known that the graph $K_{1,6}$ is not realizable as a unit disk graph.  We first show that $K_{1,6}$ is not realizable as the hypergraph of a wireless network.

\begin{Proposition} \label{prop:K16:not:realizable}
For all values of the path loss exponent $\gamma > 0$, the hypergraph $K_{1,6}$ is not realizable. 
\end{Proposition}

\noindent \emph{Proof:}
Fix $\gamma > 0$.  By way of contradiction, suppose the hypergraph $H=K_{1,6}$ is generated by some wireless network $\langle S, \gamma, \beta \rangle$.  By Lemma~\ref{lem:wlog:beta:equals:1}, it can be assumed without loss of generality that $\beta=1$. Let $S = \{s_0,s_1,\ldots,s_6\}$ be a collection of points in $\mathbb{R}^2$ such that the edge set $\mathcal{E}(H)$ is exactly $\{ \{s_0,s_i\}: i=1,\ldots,6\}$.  In particular, $W = \{s_1,\ldots,s_6\}$ is an independent set.

Translation in the Euclidean plane preserves distances, so it can be assumed without loss of generality that $s_0$ is at the origin.  For $s_i \in W$, $\{s_0,s_i\}$ is a forbidden set, whence $\frac{1}{d(s_0,s_i)^\gamma} \ge 1$, or equivalently, $d(s_0,s_i) \le 1$.  Thus, the points in $W$ lie inside the (closed) unit ball. Because $W$ is an independent set, the distance between any two points in $W$ is greater than $1$. By Lemma~\ref{lem:radially:outward}, the points in $W$ can be pushed radially outward to the unit circle, and $W$ would remain an independent set.

Thus, there exist six points $W$ on the unit circle that form an independent set.  Some two distinct points $s_i,s_j \in W$ lie on a $60^\circ$ sector of the unit circle, and the distance between these two points is at most $1$.  It follows that $\{s_i,s_j\}$ is a forbidden set, a contradiction.
\qed

\begin{Definition}
Let $r(\gamma)$ be a function from the domain $(0,\infty)$ to the positive integers, defined as follows: 
$r(\gamma):=$ the largest integer $k$ such that the hypergraph $K_{1,k}$ is generated by some wireless network with path loss exponent $\gamma$.
\end{Definition}

In other words, $r(\gamma)$ is the maximal value of $k$ such that the star hypergraph $K_{1,k}$ is realizable for the given value of $\gamma$.   In Corollary~\ref{cor:max:r:K1_r:realizable}, it was shown that when $\gamma=4$, the hypergraph $K_{1,4}$ is realizable and the hypergraph $K_{1,5}$ is not realizable. This implies that $r(4)=4$.  The results presented in this paper also indicate that $r(3)=4$, $r(2)=3$,  and $r(\gamma)=2$ for $\gamma \in (0,1]$. Additionally, Proposition~\ref{prop:K16:not:realizable} establishes that  $r(\gamma) \le 5$ for $\gamma > 0$.  These findings are summarized in Table~\ref{tab:r_alpha_values}. 

\begin{table}[htbp]
	\centering
	\begin{tabular}{c|ccccc}
		$\gamma$ & $(0,1]$ & $2$ & $3$ & $4$ & $[5,\infty)$ \\
		\hline
		$r(\gamma)$ & $2$ & $3$ & $4$ & $4$ & $\le 5$ \\
	\end{tabular}
	\caption{Values of $r(\gamma)$ known so far.}
	\label{tab:r_alpha_values}
\end{table}

Observe that as $\gamma$ increases, $r(\gamma)$ remains constant or increases. Intuitively, this behavior is expected because, with a higher path loss exponent $\gamma$, the signal strength diminishes more rapidly with distance, thereby reducing interference. This reduction in interference can potentially enable more stations to transmit simultaneously. Thus, we intuitively expect $r(\gamma)$ to be a nondecreasing function of $\gamma$. This is formally proved next. We begin by proving a stronger statement, assuming a reception threshold of $\beta = 1$, which is sufficient to establish the desired result.

\begin{Lemma} \label{lem:relationship:amplified}
Let $\gamma' > \gamma > 0$.  If the wireless network $\langle S, \gamma, 1 \rangle$ generates the hypergraph $K_{1,k}$, then the wireless network $\langle S, \gamma', 1 \rangle$ generates the hypergraph $K_{1,k}$ for all $\gamma' > \gamma$. 
\end{Lemma}

\noindent \emph{Proof:}
Suppose the wireless network $\langle S, \gamma, 1 \rangle$ generates the hypergraph $K_{1,k}$. In this hypergraph, adjacent nodes must be within a distance of at most 1, while nonadjacent nodes are separated by a distance greater than 1. Increasing the path loss exponent $\gamma$, which raises distance values to a higher power, only amplifies this distance-based relationship. Hence, the assertion follows.
\qed

\begin{Proposition} \label{prop:r:nondecreasing}
	The function $r(\gamma)$ is a nondecreasing function of $\gamma$.
\end{Proposition}

\noindent \emph{Proof:}
Fix $\gamma > 0$.  Suppose the hypergraph $K_{1,k}$ is generated by some wireless network having path loss exponent $\gamma$.  It suffices to show that for each $\gamma' > \gamma$,  there exists a wireless network with path loss exponent $\gamma'$ that generates $K_{1,k}$.

Fix $\gamma'$ to be any value greater than $\gamma$. By hypothesis, there exists a wireless network $\mathcal{A} = \langle S, \gamma, \beta \rangle$ generating $H=K_{1,k}$.  By Lemma~\ref{lem:wlog:beta:equals:1}, it can be assumed without loss of generality that $\beta=1$.  By Lemma~\ref{lem:relationship:amplified}, the hypergraph generated by the wireless network $\langle S, \gamma', 1 \rangle$ is the hypergraph $K_{1,k}$.
\qed

\begin{Theorem}\label{thm:K15:alpha:ge:5}
For all values of the path loss exponent $\gamma \ge 5$, the hypergraph $K_{1,5}$ is realizable. 
\end{Theorem}

\noindent \emph{Proof:}
By Proposition~\ref{prop:K16:not:realizable} and Proposition~\ref{prop:r:nondecreasing}, in order to prove the realizability of $K_{1,5}$ for all $\gamma \ge 5$, it suffices to prove the realizability of $K_{1,5}$ for $\gamma=5$.  So let $\gamma=5$.

Consider the wireless network $\mathcal{A} = \langle S, \gamma, \beta \rangle$, where $S = \{s_0,s_1,\ldots,s_5\}$, $\gamma=5$, $\beta=1$, $s_0$ is at the origin, and $W = \{s_1,\ldots,s_5\}$ is a set of $5$ points placed uniformly on the unit circle in counterclockwise order.  Let $s_1$ have polar coordinates $(1,0)$, and more generally, let $s_i$ have polar coordinates $(1,(i-1) \frac{2 \pi}{5})$, for $i \in [5]$.  Since $d(s_0,s_i)=1$, the subsets $\{s_0,s_i\}$ are forbidden sets $(i \in [5])$. To show that the hypergraph generated by the wireless network $\mathcal{A}$ is $K_{1,5}$, it suffices to show that $W$ is an independent set, i.e. that
$$ E(W-\{s_1\},s_1) = \sum_{i=2}^{5} \frac{1}{d(s_i,s_1)^5} < 1.$$
Here, due to symmetry, $s_1$ was chosen as the receiver without loss of generality.
The proof of this inequality is given in Lemma~\ref{lem:K15:realizable:alpha:ge:5}. 
\qed

\begin{Lemma} \label{lem:K15:realizable:alpha:ge:5}
	Let $W = \{s_1,\ldots,s_5\}$ be a set of $5$ stations placed uniformly on the unit circle. Thus,  the polar coordinates of station $s_i$ are $(1, (i-1) \frac{2 \pi}{5})$, for $i \in [5]$. Suppose the path loss exponent $\gamma=5$. Then, the energy $E(W-\{s_1\},s_1)$ at $s_1$ due to transmissions by the remaining stations satisfies the inequality
$$ E(W-\{s_1\},s_1) = \sum_{i=2}^{5} \frac{1}{d(s_i,s_1)^5} < 1.$$
\end{Lemma}

\noindent \emph{Proof:}
The proof is straightforward. The rectangular coordinates of $s_1$ are $(1,0)$. From elementary trigonometry, we have that that the rectangular coordinates of $s_2$ and $s_3$ are given by
$$s_2 = \left(\cos \frac{2 \pi}{5}, \sin \frac{2 \pi}{5} \right) = \left( \frac{-1+\sqrt{5}}{4}, \frac{\sqrt{10+2\sqrt{5}}}{4} \right)$$
and
$$s_3 = \left(\cos \frac{4 \pi}{5}, \sin \frac{4 \pi}{5} \right) = \left( \frac{-1-\sqrt{5}}{4}, \frac{\sqrt{10+2\sqrt{5}}(-1+\sqrt{5}) }{8} \right).$$
By symmetry, $d(s_4,s_1)=d(s_3,s_1)$ and $d(s_5,s_1)=d(s_2,s_1)$, whence
$$ E(W-\{s_1,\},s_1)  =  \sum_{i=2}^{5} \frac{1}{d(s_i,s_1)} = \frac{2}{d(s_2,s_1)^5} + \frac{2}{d(s_3,s_1)^5}.$$
Applying the Pythagorean theorem and simplifying, one obtains the distance values 
$$d(s_2,s_1) = \frac{\sqrt{10-2 \sqrt{5}}}{2}, d(s_3,s_1)=\frac{\sqrt{10+2 \sqrt{5}}}{2}.$$
Raising their reciprocals to the fifth power and simplifying, one gets that
$$E(W-\{s_1,\},s_1) = \sqrt{\frac{100 + 8 \sqrt{5}}{125}},$$
which is less than $1$.  
\qed

Proposition~\ref{prop:K16:not:realizable} and Theorem~\ref{thm:K15:alpha:ge:5} imply the following:

\begin{Corollary}
We have that $r(\gamma)=5$, for all $\gamma \ge 5$.
\end{Corollary}

So far, analytical methods were used to obtain the value of $r(\gamma)$ for certain values of $\gamma$.  We now conduct computer simulations to extend these results.  To this end, a special case - where the stations are placed uniformly on the unit circle - will first be investigated. As we shall see, the values obtained from these simulations provide a tight lower bound for and usually the exact value of $r(\gamma)$.

Let $\mathcal{U}_k$ denote a set $\{s_1,s_2,\ldots,s_k\}$ of $k$ stations $(k \ge 1)$ placed uniformly on the unit circle, where the polar form of the location of $s_i$ is $s_i=\left(1, (i-1) \frac{2 \pi}{k}\right)$, for $i \in [k]$. Recall that $\mathcal{U}_k$ is feasible for some wireless network with path loss exponent $\gamma$ iff 
$$E(\mathcal{U}_k - \{s_1\},s_1) = \sum_{i=2}^{k} \frac{1}{d(s_i,s_1)^\gamma} < 1,$$
where by symmetry $s_1$ was chosen as the receiver without loss of generality, and where by Lemma~\ref{lem:wlog:beta:equals:1} the reception threshold was chosen to be $1$ without loss of generality.

Let $u(\gamma)$ be a function from the domain $(0,\infty)$ to the positive integers, defined as follows: $u(\gamma)$ is defined to be the largest integer $k$ such that the set $\mathcal{U}_k$ is feasible for some wireless network with path loss exponent $\gamma$. Thus, $u(5)=5$ by Lemma~\ref{lem:K15:realizable:alpha:ge:5} and Proposition~\ref{prop:K16:not:realizable}.  

\begin{Lemma}\label{u:alpha:nondecr}
The function $u(\gamma)$ is a nondecreasing function of $\gamma$. 
\end{Lemma}

\noindent \emph{Proof:}
Fix $\gamma, \gamma',$ with $\gamma < \gamma'$.  Suppose $\mathcal{U}_k$ is feasible for some wireless network with path loss exponent $\gamma$ and  with reception threshold $\beta=1$. Then, for path loss exponent $\gamma'$, we have
$$E(\mathcal{U}_k - \{s_1\}, s_1) = \sum_{i=2}^k \frac{1}{d(s_i,s_1)^{\gamma'}}.$$

The Euclidean distance between any two distinct points in $\mathcal{U}_k$ is greater than $1$.  For all $s_i \in \mathcal{U}_k - \{s_1\}$,  because $d(s_i,s_1) > 1$, it follows that $d(s_i,s_1)^{\gamma'} > d(s_i,s_1)^\gamma$. Thus,
$$\sum_{s_i \in \mathcal{U}_k -\{s_1\}}\frac{1}{d(s_i,s_1)^{\gamma'}}  
< \sum_{s_i \in \mathcal{U}_k -\{s_1\}}\frac{1}{d(s_i,s_1)^\gamma} 
< 1.$$
It follows that $\mathcal{U}_k$ is an independent set for the wireless network $\langle \mathcal{U}_k, \gamma', 1 \rangle$. 
\qed

Because $u(\gamma)$ is a nondecreasing function of $\gamma$, the bisection search algorithm can be used to compute the smallest value of $\gamma$ such that $\mathcal{U}_k$ is feasible for some wireless network with path loss exponent $\gamma$, as shown in Algorithm~\ref{algo:bisection:smallest:alpha}.  In this algorithm, the main function \texttt{smallestGammaUkIsFeasible} takes as arguments a positive integer $k$ and an initial search space $[low,high]$ for bisection search, taken to be $[1,10]$ in our instances.  This function makes calls to the boolean function \texttt{isUkFeasible(gamma,k)}, which returns True iff $\sum_{i=2}^{k} \frac{1}{d(s_i,s_1)^\gamma} < 1$, where the $s_i$'s are points spaced uniformly on the unit circle.  When the main function is called with parameters $(3,1,10)$, $(4,1,10)$, and $(5,1,10)$, the values returned  are 
1.2618565559387207, 2.543102741241455, and 4.855621814727783, respectively.  In the rest of this exposition, for simplicity we shall round these numerical values to three decimal places.  The code used to generate these simulation results is available on GitHub at \cite{ashwin:ganesan:GitHub:repo}, allowing for easy reproduction and extension of the results.

\begin{algorithm}
\DontPrintSemicolon
\KwIn{A 3-tuple, consisting of a positive integer $k$ and an initial search space $[\mbox{low},\mbox{high}]$ for bisection search}
\KwOut{The smallest $\gamma$ such that $\mathcal{U}_k$ is feasible for the wireless network $\langle \mathcal{U}_k, \gamma,1 \rangle$}
\SetKwBlock{Begin}{function}{end function}
\Begin($\text{smallestGammaUkIsFeasible} {(} k,low,high {)}$)
{
  mid = (low + high) / 2\;
  tolerance = 0.00001\;
  \While{$|\mbox{high} - \mbox{low}| > \mbox{tolerance}$}
  {
   \uIf{isUkFeasible(mid,k)}
   {
     high = mid\;
   }
   \Else
   {
   low = mid\;
   }
   mid = (low + high) / 2\;
  }
  \Return{mid}

}
\caption{Bisection search to determine smallest value of the path loss exponent $\gamma$ such that $\mathcal{U}_k$ is feasible}\label{algo:bisection:smallest:alpha}
\end{algorithm}

The function $u(\gamma)$ is a nondecreasing, integer-valued function and increases like the unit step function. Computer simulations using the bisection method give that $u(\gamma)$ increases from $2$ to $3$ when $\gamma$ is approximately $1.262$, increases from $3$ to $4$ when $\gamma$ is approximately $2.543$, and increases from $4$ to $5$ when $\gamma$ is approximately $4.856$.  It is clear that $\mathcal{U}_2$ is feasible for all $\gamma > 0$.  To summarize, we have that
$$u(\gamma) = \begin{cases}
2, ~~~\mbox{if } \gamma \in (0,1.262) \\
3, ~~~\mbox{if } \gamma \in (1.262, 2.543) \\
4, ~~~\mbox{if } \gamma \in (2.543, 4.856) \\
5, ~~~\mbox{if } \gamma  > 4.856 
\end{cases}
$$

The transition value of $1.262$ obtained above using computer simulations can also be derived using analytical methods:

\begin{Lemma}
Consider the wireless network $\langle \mathcal{U}_3, \gamma, 1 \rangle$, where $\mathcal{U}_3 = \{s_1, s_2, s_3\}$ is a set of three stations placed uniformly on the unit circle, $\gamma$ denotes the path loss exponent, and the reception threshold is $1$.  Then, $\mathcal{U}_3$ is feasible if and only if $\gamma > 2 \log_3 2 \approx 1.262$. 
\end{Lemma}

\noindent \emph{Proof:}
Because the three stations are placed uniformly on the unit circle, the distance between any two distinct stations is $\sqrt{3}$.  Choose a station from $\mathcal{U}_3$, say $s_1$.  The interference at $s_1$ due to transmissions by $s_2$ and $s_3$ is
$$E(\{s_2,s_3\}, s_1) = \frac{1}{(\sqrt{3})^\gamma} + \frac{1}{(\sqrt{3})^\gamma} = \frac{2}{(\sqrt{3})^\gamma},$$
which is less than $1$ if and only if $\gamma > 2 \log_3 2$. 
\qed

The connection between $u(\gamma)$ and the function of interest, namely $r(\gamma)$, is made next. 

\begin{Lemma} \label{lem:u:le:r}
We have that $u(\gamma) \le r(\gamma)$, for all $\gamma > 0$. 
\end{Lemma}

\noindent \emph{Proof:}
Fix $\gamma > 0$ and let $k$ be any positive integer.  Suppose $\mathcal{U}_k$ is feasible for some wireless network with path loss exponent $\gamma$.  By Lemma~\ref{lem:wlog:beta:equals:1}, without loss of generality it can be assumed that this wireless network has reception threshold $\beta=1$.  It suffices to show that $r(\gamma) \ge k$.  Let $S = \mathcal{U}_k \cup \{s_0\}$, where $s_0$ is at the origin.  The hypergraph generated by the wireless network $\langle S, \gamma, 1 \rangle$ is the $2$-uniform hypergraph $K_{1,k}$.  It follows that $r(\gamma) \ge k$. 
\qed

The value of $r(\gamma)$ can now be obtained for various nonintegral values of $\gamma$.  For $\gamma$ in the interval, $(1.262,2]$,  $u(\gamma)=3$; by Lemma~\ref{lem:u:le:r}, $r(\gamma) \ge 3$.  Furthermore, $r(2)=3$. Since $r(\gamma)$ is a nondecreasing function, $r(\gamma) \le 3$, for all $\gamma \in (1.262,2]$.  Together, these results imply that $r(\gamma)=3$ for all $\gamma \in (1.262,2]$.  Similarly, one can put together the results on $u(\gamma)$ obtained above using computer simulations to determine $r(\gamma)$ for various nonintegral values of $\gamma$. We then obtain:

\begin{Corollary} \label{cor:r:alpha:nontegral} We have that
$$r(\gamma) = \begin{cases}
2, ~~~\mbox{if } \gamma \in (0, 1] \\
3, ~~~\mbox{if } \gamma \in (1.262, 2] \\
4, ~~~\mbox{if } \gamma \in (2.543, 4] \\
5, ~~~\mbox{if } \gamma  > 4.858 
\end{cases}
$$
\end{Corollary}

We leave for further research the problems of determining $r(\gamma)$ in the intervals $(1, 1.262)$, $(2,2.543)$ and $(4, 4.858)$, the question of whether equality holds in Lemma~\ref{lem:u:le:r}, and the problems of obtaining the transition values $2.543$ and $4.858$ exactly using analytical methods. 

\section{Hypergraphs Of Line Networks}
\label{sec:line:networks}

In this section, we investigate line networks, which are wireless networks $\langle S, \gamma, \beta \rangle $ where the set $S$ of stations lies on a straight line such as the $x$-axis.  Line networks are an interesting and simple special case to study first and have been studied by many researchers \cite{Kose:Medard:PIMRC:2017} \cite{Ganesan:ToN:2020}.  We assume throughout that the path loss exponent $\gamma=4$.

The \emph{path graph} on $n$ vertices $(n \ge 1)$, denoted by $P_n$, is the graph with vertex set $\{1,2,\ldots,n\}$ and with edge set $\{ \{i, i+1\}: i = 1, \ldots, n-1\}$ if $n \ge 2$ and with the empty edge set if $n=1$.  The \emph{path hypergraph} $P_n$ $(n \ge 1)$ is the hypergraph consisting of the same vertex set and edge set as the path graph $P_n$. 

To motivate the problem studied in this section, first consider the simpler model of unit disk graphs.  Let $S=\{s_i: i=1,\ldots,n \}$ be a set of stations, where the locations of the stations are given by $s_i=(i,0) (i=1,\ldots,n)$, so that the distance between any two consecutive stations is exactly $1$ unit.  Under the unit disk graph model, the conflict graph would be the path graph $P_n$ on $n$ vertices.  Hence, the path graph $P_n$ is a unit disk graph, and we would like to generalize this result.  A natural question that arises is the following: Does the hypergraph $P_n$ arise as the interference model of a wireless network?  In other words, is the path hypergraph $P_n$ realizable?  If $S$ is assumed to be the set of $n$ locations above, then every pair of consecutive vertices is forbidden, and so $P_n$ is contained in the hypergraph generated by the wireless network $\langle S, \gamma, \beta \rangle$.  The question is whether the hypergraph contains other minimal forbidden subsets besides the edge set of $P_n$.  In other words,  can the cumulative interference from multiple nodes, no two of which are consecutive, be intolerable?  We answer this question in the negative, thereby proving that the hypergraph $P_n$ is realizable.

\begin{Theorem} \label{thm:hyp:is:P_n}
Suppose the path loss exponent $\gamma=4$. Then, the path hypergraph $P_n$ is realizable for all $n \ge 1$. 
\end{Theorem}

\noindent \emph{Proof:} 
A wireless network $\langle S, \gamma, \beta \rangle$ will be given that generates the path hypergraph $P_n$.  The assertion clearly holds if $n=1$, so suppose $n \ge 2$. Without loss of generality, it can be assumed that the reception threshold $\beta=1$.  Recall that $\gamma=4$.  Let $S = \{s_1,s_2,\ldots,s_n\}$, where $s_i$ is at location $(i,0)$ for all $i$.  Thus, the spacing between two consecutive stations on the line is exactly $1$.  Let $H=(S, \mathcal{E})$ denote the hypergraph generated by the wireless network $\langle S, \gamma, \beta \rangle$.  Because the distance between any two consecutive stations on the line is at most $1$ and the distance between any two nonconsecutive stations on the line is greater than $1$, the minimal forbidden sets of size $2$ are exactly the edges of the path graph $P_n$.  In other words, $\mathcal{E}$ contains the edge set of $P_n$. It  suffices to show that there are no other minimal forbidden sets.

Let $W \subseteq S$ be a subset of size at least $3$.  If $W$ contains any edge of $P_n$, then $W$ would be a forbidden set but not minimal and so $W \notin \mathcal{E}$.   So suppose $W$ is a set of pairwise nonconsecutive stations.  In other words, suppose $W$ satisfies the property that the distance between any two distinct elements of $W$ is at least $2$.  It suffices to show that $W$ is feasible, for this would imply that $\mathcal{E}$ contains no further elements than the edge set of $P_n$. Fix $s_i \in W$.  It suffices to show that $E(W-\{s_i\}, s_i) < 1$. 

We have that
\begin{eqnarray*}
E(W-\{s_i\}, s_i) & = & \sum_{s_j \in W - \{s_i\}} \frac{1}{d(s_j,s_i)^4}  \\
& < & 2 \sum_{j=1}^n \frac{1}{(2j)^4}  \\ 
& < & 2 \sum_{j=1}^\infty \frac{1}{(2j)^4}   \\
& = & \frac{1}{8} \sum_{j=1}^\infty \frac{1}{j^4}   \\
& = & \frac{1}{8} \frac{\pi^4}{90}   \\
& < & 1.  \\
\end{eqnarray*}

The first inequality above follows from the fact that for any station $s_i \in W$ whose location is $(i,0)$, the two stations in $W-\{s_i\}$ closest to $s_i$ would be at least as far apart from $s_i$ as $s_{i-2}$ and $s_{i+2}$ are, since the stations in $W$ are pairwise nonconsecutive.  The next two stations in $W-\{s_i\}$ closest to $s_i$ would be $s_{i-4}$ and $s_{i+4}$ (or stations even further away from $s_i$ than these two stations).   There are at most $n$ stations on either side of $s_i$.  Thus, the interference at $s_i$ due to transmissions by the remaining stations in $W$ can be upper bounded as shown by the first inequality. The second inequality is clear because we are replacing the finite sum by an infinite series. The last equality follows from Euler's formula that the sum of the reciprocals of the fourth powers of the positive integers is given by $\sum_{n=1}^\infty \frac{1}{n^4} = \frac{\pi^4}{90}$ \cite{Euler:1740:Latin} \cite{Euler:1740:English}.  It follows that the hypergraph generated by the wireless network $\langle S, \gamma, \beta \rangle$ is the path hypergraph $P_n$. 
\qed

Theorem~\ref{thm:hyp:is:P_n}, which states that the path hypergraph $P_n$ is realizable, assumed that the path loss exponent $\gamma=4$.  If the path loss exponent $\gamma=2$, the assertion still holds.  The same wireless network realization given in the proof of Theorem~\ref{thm:hyp:is:P_n} can be used along with Euler's formula $\sum_{n=1}^\infty \frac{1}{n^2} = \frac{\pi^2}{6}$ to prove this assertion.

\section{Concluding Remarks} \label{sec:concluding:remarks}

In the present work, we investigated the properties of a hypergraph invariant - the interference degree of a hypergraph. The problem of computing the interference degree of a  hypergraph was shown to be NP-hard and some  properties of the interference degree of hypergraphs were proved.   Furthermore, we investigated the properties of hypergraphs that are realizable, i.e. that arise in practice due to physical (SINR) constraints.  It is believed that the additional structure possessed by realizable hypergraphs can be exploited for efficient algorithm design.

An open problem in the literature is to extend the results in the literature on unit disk graphs to hypergraph models.  It is known that the maximal value of $r$ such that the graph $K_{1,r}$ is a unit disk graph is $r=5$.  In the present work, these results were extended to hypergraph models. More specifically, the function $r(\gamma)$ was defined to be the maximal value of $k$ such that the $2$-uniform hypergraph $K_{1,k}$ is generated by some wireless network with  path loss exponent  $\gamma$.  We obtained the exact value of $r(\gamma)$ for various integer values of $\gamma$. These results were further extended to nonintegral values of the path loss exponent $\gamma$. We also investigated hypergraphs generated by line networks and showed that the path graph $P_n$ is realizable when $\gamma=4$. 

There are several directions for further research. Corollary~\ref{cor:r:alpha:nontegral} gives the value of $r(\gamma)$ for certain intervals.  Firstly, determining $r(\gamma)$ for the remaining values of the path loss exponent $\gamma$ and obtaining the transition values in Corollary~\ref{cor:r:alpha:nontegral} exactly using analytical methods are directions for further research. Secondly, determining which other structured hypergraphs besides the stars $K_{1,k}$ arise in practice as the interference model of wireless networks, based on physical constraints, is an open problem.  Finally, it would be worth investigating the maximum possible interference degree of realizable hypergraphs, as this gives a performance guarantee on distributed algorithms for wireless networks under the hypergraph interference model.  

\section{Acknowledgements}

Thanks are due to the anonymous reviewers for helpful comments and suggestions.

 {
\bibliographystyle{myplain}
\bibliography{refs_ag.bib}

}
\end{document}